\documentclass[%
 reprint,
 superscriptaddress,
 amsmath,amssymb,
 aps,
 pra,
 floatfix,
 twocolumn,
]{revtex4-2}

\newcommand{\ket}[1]{\lvert #1\rangle}

\usepackage{graphicx}
\usepackage{dcolumn}
\usepackage{bm}
\usepackage{hyperref}

\usepackage{tikz}
\usepackage{dsfont}
\usetikzlibrary{calc}
\usepackage{comment}

\usepackage{qcircuit}

\usepackage{orcidlink}

\usepackage{amsmath}
\usepackage{upgreek}
\usepackage{physics}
\usepackage{isomath}
\usepackage{placeins}
\usepackage{float}
\begin{document}


\title{Universal Gate Set for Optical Lattice Based Atom Interferometry}

\author{Catie LeDesma}
\email[]{Catherine.LeDesma@colorado.edu}
\affiliation{JILA \& Department of Physics, University of Colorado Boulder, CO 80309-0440, USA}

\author{Kendall Mehling}
\affiliation{JILA \& Department of Physics, University of Colorado Boulder, CO 80309-0440, USA}

\author{John Drew Wilson}
\affiliation{JILA \& Department of Physics, University of Colorado Boulder, CO 80309-0440, USA}

\author{Marco Nicotra}

\affiliation{College of Engineering and Applied Science, University of Colorado Boulder, Boulder, Colorado 80309, USA}

\author{Murray Holland}
\affiliation{JILA \& Department of Physics, University of Colorado Boulder, CO 80309-0440, USA}


\date{\today}

\begin{abstract}
In this paper, we propose a new paradigm for atom interferometry and demonstrate that there exists a universal set of atom optic components for inertial sensing. These components constitute gates with which we carry out quantum operations and represent input-output matterwave transformations between lattice eigenstates. Each gate is associated with a modulation pattern of the position of the optical lattice according to machine-designed protocols. In this methodology, a sensor can be reprogrammed to respond to an evolving set of design priorities without modifying the hardware. We assert that such a gate set is metrologically universal, in analogy to universal gate sets for quantum computing. Experimental confirmation of the designed operation is demonstrated via in situ imaging of the spatial evolution of a Bose-Einstein condensate in an optical lattice, and by measurement of the momentum probabilities following time-of-flight expansion. The representation of several basic quantum sensing circuits is presented for the measurement of inertial forces, rotating reference frames, and gravity gradients.
\end{abstract}

{
\let\clearpage\relax
\maketitle
}

\vspace*{-1pc}
\section{Introduction}\label{sec:1D}
\vspace*{-.75pc}

Major advancements in quantum metrology over the past few decades have placed optical lattice clocks~\cite{Takamoto2005, PhysRevLett.130.113203} and atomic interferometers~\cite{ PhysRevLett.66.2693, PhysRevLett.67.181} as enabling technologies for fundamental physics research. One successful maturation of quantum sensing has been the ever increasing precision of matterwave sensors capable of measuring inertial forces~\cite{LightPulse_Kasevich}, rotations~\cite{PhysRevLett.78.2046, s120506331}, and gravity gradients~\cite{PhysRevLett.81.971, Bertoldi2006}. Atomic interferometers have also established constraints on fundamental constants~\cite{PhysRevLett.70.2706, PhysRevLett.100.050801}, set certain dark energy bounds~\cite{doi:10.1126/science.aaa8883}, and provided the most rigorous studies of the equivalence principle~\cite{EquivPrinc_Kasevich, Herrmann_2012}. While atomic interferometers have proven to be excellent precision measurement devices in the laboratory, many forefront objectives such as simultaneous sensing in multiple dimensions~\cite{PhysRevLett.122.043604, doi:10.1126/sciadv.aau7948} and demonstrations of practical field-deployable architectures~\cite{Ménoret2018, Lee2022} have remained an ongoing challenge.

Recent advances in atomic clocks suggest that a powerful path forward is one which makes use of optical lattices for control, and indeed recent experimental results have shown that an emergent class of atom interferometers using three-dimensional optical lattices may offer an alternative solution~\cite{ledesma2024vector}. These have been termed Bloch-Band Interferometers~(BBIs) since they accumulate the inertial phase in the Bloch band eigenstates of an optical lattice, rather than in free space as in traditional atom optic devices. The use of stationary eigenstates of the system allows for phase accumulation over long interrogation times, a necessary condition for precision metrology applications. In addition, an optical lattice can impart significant impulse force, offering the potential for a robust platform for sensing in the face of dynamically harsh environments with inherent thermal and vibrational noise. Furthermore, this system offers the opportunity for dynamic reconfigurability of the sensing functionality without the need for any modifications to the physical device itself. 

In this paper, we propose a basis for atom interferometry by exploring a universal gate set of operations that can be sequenced to build any inertial sensing device. We believe it is useful to point out the natural connection between this new design philosophy and the universal gate sets of quantum computing~\cite{Feynman1986, PhysRevA.51.1015}, in which a handful of elementary operations acting upon qubits can be applied sequentially to execute a quantum algorithm of arbitrary complexity. In the case of quantum logic gates, a finite number of qubits connect to the input and a unitary operator maps to the output by modifying the quantum amplitudes of each state according to the gate function. Here, we do the same thing, but instead of quantum logic gates, our gates are elementary input-output operations that perform unitary transformations on quantum matterwaves between lattice eigenstates~\cite{PhysRevA.64.063613, PRXQuantum.2.040303, 10015539}. The concept of universality is established in a metrological context, meaning the ability to sense potential fields in three dimensions of arbitrary functional form, including derivatives. This framework unifies a wide variety of sensors including accelerometers, magnetometers, gravimeters, gyroscopes, and gradiometers.

\begin{figure*}[t]
    \centering
    \includegraphics[scale=0.30]{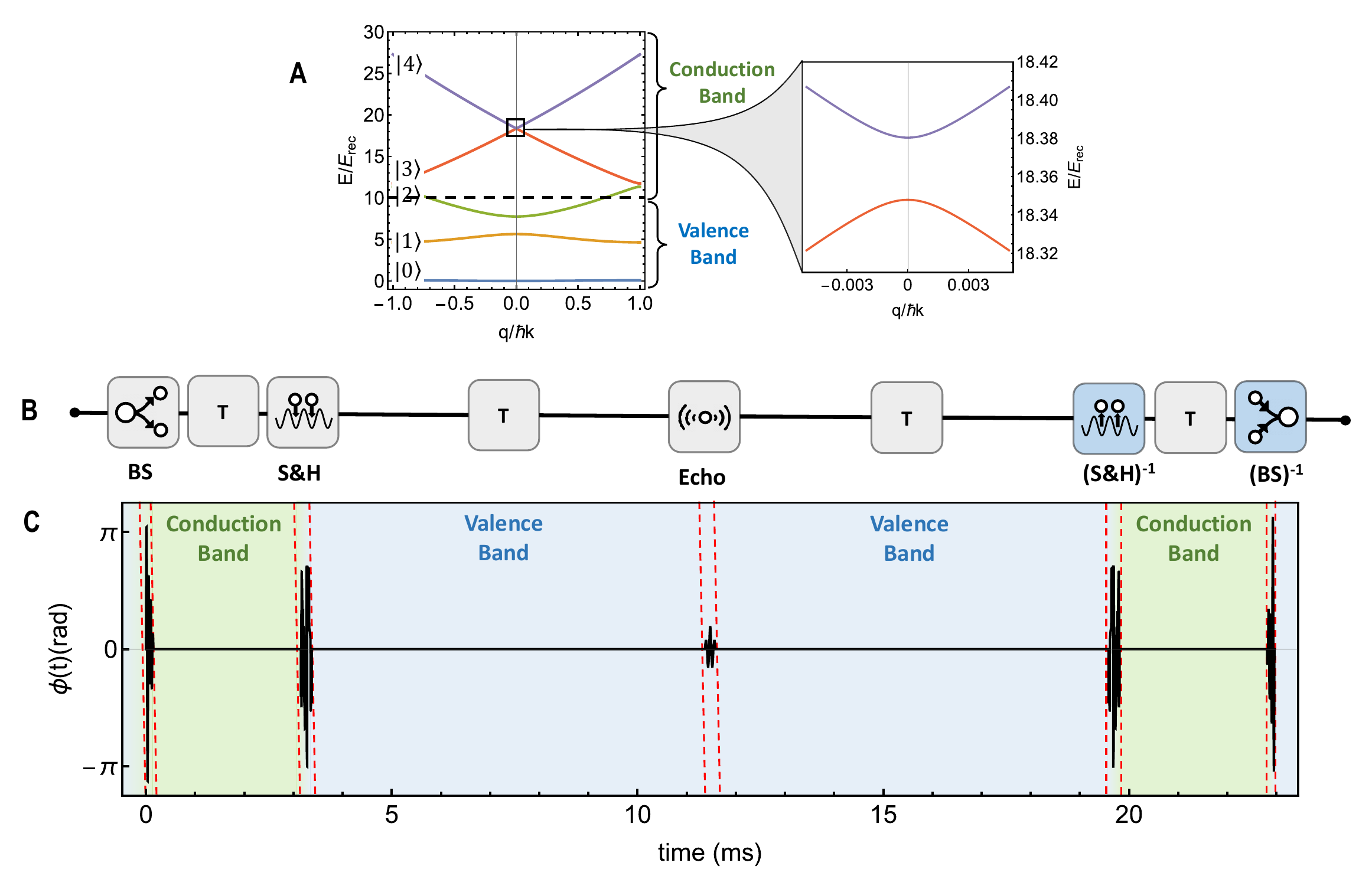}
\caption{\textbf{Bloch-band interferometry (BBI)} \textbf{A:} Band energies as a function of quasimomentum for a lattice depth $10E_r$ (dashed line), the exploded graph magnifying the small avoided crossing of the two conduction bands. \textbf{B:}~Example sequence of a BBI sensor circuit producing an accelerometer with a hold stage. The tiles are specific matterwave gates that will all be described in this paper. \textbf{C:}~The timing protocol of the control function that results, $\phi(t)$, showing the lattice shaking during the gate operations, with the $2\pi$ radian range corresponding to moving the lattice by a single optical wavelength. The lattice is fixed in space during most of the sequence, with the atoms in either the conduction or valence band eigenstates.} 
\label{fig:Bloch Bands of the First Brillouin Zone}
\end{figure*}

While some of the gates we present are familiar from other contexts, such as beamsplitters and mirrors, here we also present for the first time new atom optic components such as ``Split \& Hold" and ``Echo'' that broaden the class of potential sensors. Each gate is associated with a control function that specifies a timing protocol for the lattice position~\cite{PhysRevLett.81.5093, PhysRevLett.90.094101} and is designed using machine-learning and quantum optimal control algorithms. For each gate, we provide high-fidelity examples of these protocols and show their performance through numerical simulation. To demonstrate that they work in practice, we implement all the gates and confirm their anticipated function in our rubidium Bose-Einstein condensation experiment (BEC) using a one-dimensional optical lattice~\cite{Denschlag2002}. We do this by imaging directly the evolution of the BEC in situ, while the atoms are held in the optical lattice, and by measuring the momentum distribution after time-of-flight (TOF) expansion. Following presentation of the gate set, we explicitly show how the components can be sequenced in atom optic circuits to construct a variety of basic sensing devices and evaluate the resulting sensitivity for parameter estimation.

\vspace*{-1pc}
\subsubsection*{\bf Interferometry using the Bloch bands of an optical~lattice}
\vspace*{-.75pc}

The basic framework of BBI is sketched out in Fig.~\ref{fig:Bloch Bands of the First Brillouin Zone}. We illustrate the energy dispersion of the lowest 5 Bloch bands for a lattice of depth $10E_r$, where $E_r=\hbar^2k_L^2/2m$ is the recoil energy. The recoil energy depends on the the atomic mass $m$ and the lattice wavevector $k_L=2\pi/\lambda$ through the optical wavelength~$\lambda$. The bands are a locus of energy eigenstates, each eigenstate labeled by the band index, a non-negative integer, as well as the continuous quasimomentum that takes possible values spanning the Brillouin zone. As shown in Fig.~\ref{fig:Bloch Bands of the First Brillouin Zone}A, the bands with the lowest three indices, $\ket{0}$, $\ket{1}$, and $\ket{2}$, which we refer to as valence band states, do not possess enough energy to propagate between lattice sites without tunneling and are localized in the lattice potential wells. The conduction band states, $\ket{3}$ and $\ket{4}$, display essentially the quadratic energy dispersion of free particles indicating that the atoms are able to transit the lattice easily and therefore cover large distances. The ability to split the atomic wavefunction by large distances is essential for achieving high sensitivity in many precision metrology applications.

To preview our gate-based sensor framework, we show an example sensor protocol---namely an accelerometer with a hold stage. Although it may not be apparent at first glance, in the sequence shown in Fig.~\ref{fig:Bloch Bands of the First Brillouin Zone}B and~C, an atom wavepacket is split along two paths, held at its furthest extents, released, propagated back, and recombined to give quantum interference and parameter readout. The sequence is displayed both as an abstract sensor circuit, with each tile representing an elementary gate, and as a timing protocol of the corresponding lattice position. Later we will analyze a universal set of gates, including the ones shown here, and for each demonstrate its functionality. As is apparent in this example, the gate transformations are generated by rapidly varying control functions of the lattice position that are produced by optimization algorithms~\cite{RLMatter_Chih}. The gate operations occupy only a small portion of the total sequence time. In between gates, the lattice position is held stationary and the atoms are in Bloch states; either in the conduction band states $\ket{3}$ and $\ket{4}$ where they propagate, or in the valence band states $\ket{0}$ and $\ket{1}$ where they are held stationary.

\vspace*{-1pc}
\subsubsection*{\bf Experimental Apparatus}
\vspace*{-.75pc}

For the experiments reported here, the sequence begins with the production of a rubidium-87 Bose-Einstein condensate (BEC) which is produced all-optically via forced evaporation in a crossed-dipole trap (CDT). A detailed description of the experimental system can be found in Refs.~\cite{ledesma2023machinedesignedopticallatticeatom, ledesma2024vectoratomaccelerometryoptical}. For the results presented in this paper we have chosen to produce condensates of order $4\times10^4$ atoms at 10~nK. Although we can create condensates that are many times larger, we intentionally do not do so to simplify interaction effects. After forming the condensate in the CDT, the BEC is adiabatically loaded into a one-dimensional optical lattice aligned in the vertical direction. The loading is performed by linearly ramping up the lattice laser intensity in a total time ranging from $400\,\mu$s to 1~ms. The lattice beams are generated using a 30~W 1064~nm fiber laser. In the same apparatus, we can perform atom interferometry experiments with three-dimensional lattices, but for the gate protocols demonstrated here, which are all one dimensional, only two counterpropagating (i.e., not retroreflected) laser beams on a single axis are used. 

Each lattice beam passes through its own independent acousto-optic modulator (AOM) that is utilized for active control of the intensity and phase of the light. The gate protocols are applied to the atoms by updating the radio frequency input sent to the lattice AOMs. We sample the theoretically calculated protocols at 50~ns resolution and experimentally create the signals by imprinting the sampled waveform onto a 80~MHz carrier, as needed for the AOM drive, using an arbitrary waveform generator~(AWG). Lattice depths were calibrated using Kapitza-Dirac diffraction~\cite{Kapitza_Dirac_1933} or by observation of the highest fidelity of applied protocols. 

We determine the momentum transformation that results from the applied gate protocol sequence through absorption imaging. This means that we expose the atoms after evolution to a resonant 780 nm laser beam directed orthogonal to the lattice direction and image the shadow of the cloud on a CMOS camera. The pixel values on the camera correspond to the integrated column density (optical density) through the quantum gas at each location. In this paper, we perform absorption imaging in two ways; both in situ imaging of atoms in the optical lattice which yields the position distribution probabilities, as well as imaging after time of flight (TOF) expansion of 10--15~ms which reveals the momentum distribution probabilities following applied operations. Successful adiabatic loading into the lattice is confirmed by observing the diffracted momentum components following TOF and comparing with what is expected for the ground state~$\ket{0}$ at the ascribed lattice depth. Since the lattice photons can only transfer quantized momentum, observed diffraction orders are easily discriminated after TOF and are separated by $2\hbar k$ along the vertical axis. 

\vspace*{-1pc}
\section{Elementary Gates}
\vspace*{-.75pc}

The one dimensional optical lattice potential, $V(x, t)$,  at each point in space $x$ and time $t$ is given by;
\begin{equation}
    V(x,t) = \frac{V_0}{2}\cos\bigl(2k_Lx + \phi(t)\bigr)
\label{eq:Lattice Potential}
\end{equation}
where $V_0$ is the lattice depth (i.e., the potential difference between intensity minima and maxima), and $\phi(t)$ is the control function. In the theoretical optimization process all gates were designed for a lattice depth of $V_0 = 10E_{\rm rec}$, that is, the band structure in Fig.~\ref{fig:Bloch Bands of the First Brillouin Zone}A. Interestingly, for any given gate, we have typically been able to find numerous control functions that look qualitatively very different but produce similar functionality when implemented in both numerical simulation and in experiment. We have used a variety of optimization methods to produce the control functions, including quantum optimal control (QPRONTO)~\cite{PhysRevA.105.032605, PhysRevA.109.012609} and machine learning (reinforcement learning)~\cite{liang2022thesis,ledesma2023machinedesignedopticallatticeatom}. 
\begin{figure}[b]
    \centering
    \hspace*{-0.5pc}
    \includegraphics[scale=0.285]{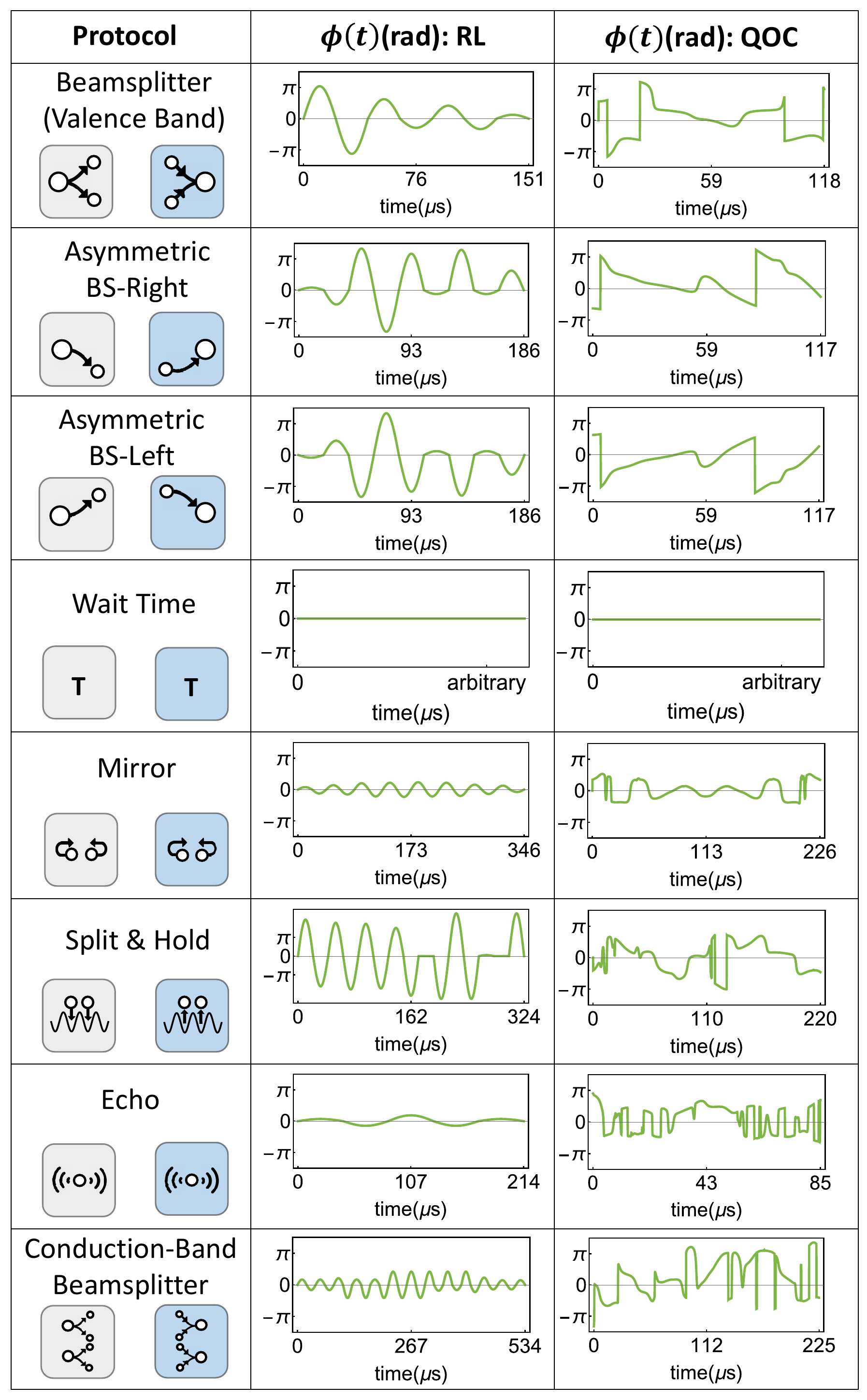}
    \caption{\textbf{Universal gate set for atom interferometry:} For each protocol the symbol (gray) as well as its reverse (blue) is shown. The waveforms are some examples of control functions for lattice shaking protocols as produced by both reinforcement learning and by quantum optimal control methods, with the horizontal axis giving times in $\mu$s and $2\pi$ on the vertical axis giving a spatial shift of one lattice wavelength. Reverse waveforms are generated by playing the control backwards.}
    \label{fig:gateset}
\end{figure}
Illustrative examples of control functions are given in Fig.~\ref{fig:gateset}. These examples are not the only solutions, and a different initialization of the neural network in reinforcement learning, or a different seed function in quantum optimal control, will usually lead to a different result. However, all the protocols shown produce gates that have in excess of 90\% fidelity with respect to their anticipated function (often in excess of 99\%). For the purposes of this paper, the exact method employed to produce high fidelity, short duration (on the order of 100s of microseconds) control functions is not essential. The effective design of control functions is a topic in its own right and will not be extensively explored here~\cite{10156455}. However, we point out that the optimization process to produce the gates can incorporate a variety of additional constraints that may respond to a specific intended application. A few examples of constraints are insensitivity to noise, insensitivity to nuisance parameters such as lattice depth error~\cite{alam2024robustquantumsensingmultiparameter, Suzuki_2020}, broad dynamic range with respect to an inertial signal, and insensitivity to mean-fields and other interaction effects. 

We now describe the elementary unitaries (gates) and observe their operation directly from the spatial evolution of the BEC in the lattice. Since absorption imaging is destructive, we run multiple experiments and take snapshots of the in situ density for an array of hold times following the application of the gate protocol. The momentum transformation is revealed via TOF by imaging both before and after applying the shaking waveform.
\vspace*{-1pc}
\subsubsection*{\bf{Beamsplitter Gate}}
\vspace*{-.75pc}

The desired functionality of the beamsplitter is a unitary operation that causes a wavepacket to split and separate in a superposition of taking two alternate paths. For our situation, the input is the ground state of the lattice~$\ket{0}$ and is macroscopically occupied by the BEC. The conduction band states $\ket{3}$ and $\ket{4}$ are attractive output beamsplitter candidates since they are eigenstates comprised primarily of an equal superposition of $\pm 4\hbar k_L$ momentum components, each with 47.4\% occupation, and have sufficient kinetic energy to exceed the maximum of the optical potential allowing atoms to move through the lattice. 
\begin{figure}[htb]
    \centering
    \includegraphics[scale=0.42]{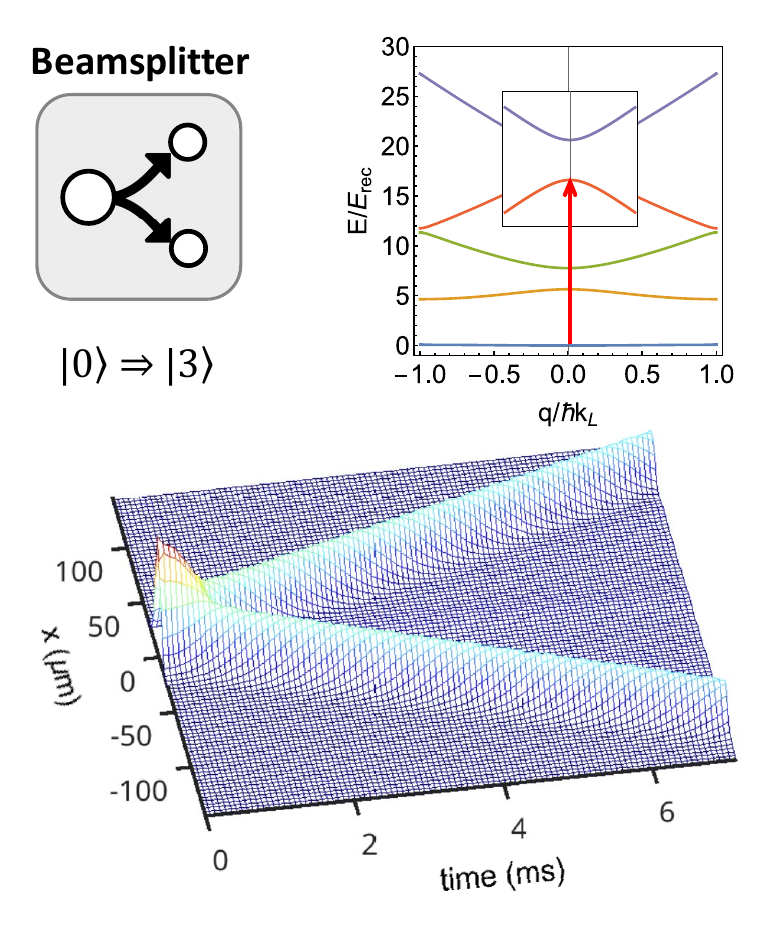}
    \vspace*{-1pc}
    \caption{\textbf{Beamsplitter design:} (Top) The beamsplitter symbol and defining Bloch state transition. (Bottom) Validation of the shaking function by numerical simulation of the Schr\"odinger evolution. The gate was first applied to the condensate and then the lattice held stationary to make apparent the splitting of the wavepacket.}
    \label{fig:bstheory}
\end{figure}
Our designated Bloch-band transformation for the beamsplitter is therefore one of two options;
\begin{equation}
\ket{0}\Rightarrow\ket{3}\qquad\mbox{or}\qquad
\ket{0}\Rightarrow\ket{4}
\end{equation}
with the first illustrated in Fig.~\ref{fig:bstheory}. The two possibilities arise from simply permuting the two output channels, a common theme for most of the components we will show. 

The high fidelity of the control function that was found for this transformation was confirmed by performing a numerical simulation. In the simulation, a quantum wavefunction was prepared in the Bloch ground state with a spatial extent of approximately 20 lattice sites. The time-dependent Schr\"odinger evolution was then solved in momentum space with the beamsplitter protocol shaking the lattice position according to Eq.~\ref{eq:Lattice Potential}. The gate protocol lasted for slightly more than 150~$\mu$s, and after application of the gate, the lattice was held stationary, and the evolution was tracked numerically for 7~ms more. The anticipated wavepacket splitting of the condensate is clearly seen in the bottom portion of Fig.~\ref{fig:bstheory}.

\begin{figure}[htb]
    \centering
    \includegraphics[scale=0.5]{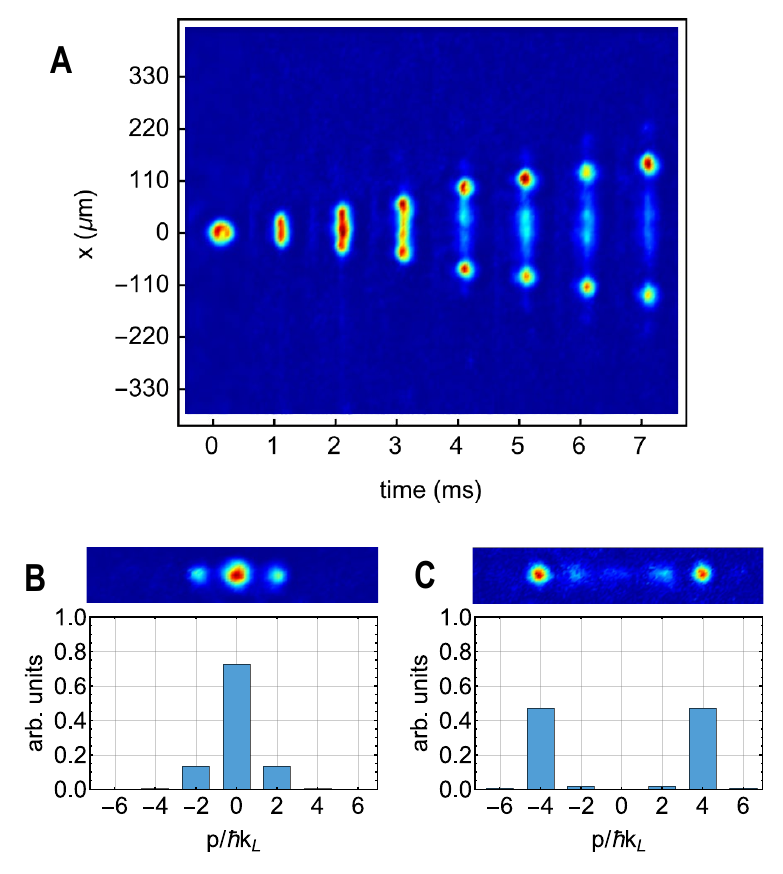}
    \caption{\textbf{Beamsplitter experiment:} {\bf (A)}  Experimental in situ images of the BEC as a function of the delay after applying the beamsplitter protocol to shake the lattice, images stacked side by side. {\bf (B)} Experimental TOF image before the beamsplitter protocol was applied. The bar graph shows the momentum probabilities from theory for the $\ket{0}$ Bloch eigenstate. {\bf (C)} The same after the beamsplitter protocol, along with the bar graph for the $\ket{3}$ target state.} 
    \label{fig:bsexpt}
\end{figure}
We validated the application of the beamsplitter protocol experimentally, and the results of this study are shown in Fig.~\ref{fig:bsexpt}. In the false color absorption images, the red regions have high atomic density and blue regions have essentially no atoms. In this paper, all false color images of this type are experimental, and are averages of 3--5 experimental shots.  When imaged, the condensate is observed to split into two clouds that linearly separate with a measured velocity commensurate with $\pm4\hbar k_L$ momenta. The in situ expansion within the lattice shows good agreement with theory and displays a separation of order 200~$\mu$m for a 7 ms hold. The momentum components that were observed by TOF before and after the gate protocol also agree well with the defining Bloch state transformation of a beamsplitter. In this paper, the TOF images showing the results of the applied protocols in momentum space (originally vertical) have all been rotated $90^{\circ}$ (shown horizontal) to enable easier comparison to the predicted theory.

\vspace*{-1pc}
\subsubsection*{\bf Asymmetric Beamsplitter Gate}
\vspace*{-.75pc}

\begin{figure}[b]
    \centering
    \includegraphics[scale=0.40]{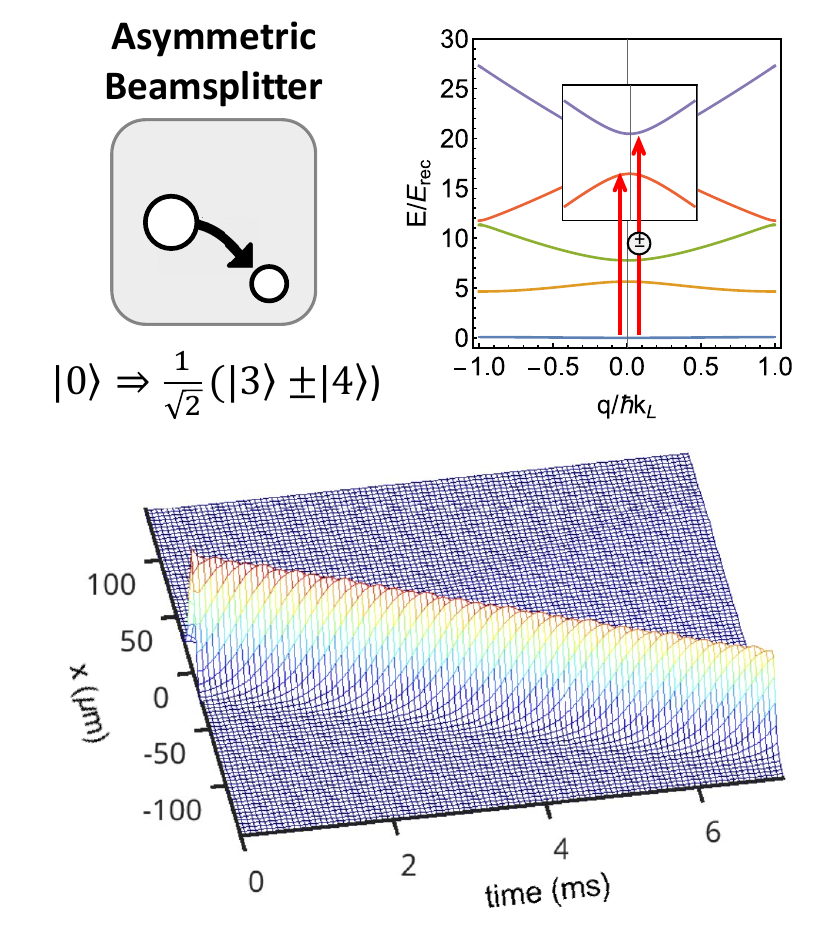}
    \vspace*{-1pc}
    \caption{\textbf{Asymmetric beamsplitter design:} (Top) As in Fig.~\ref{fig:bstheory}, but for the Bloch transition $\ket{0}\Rightarrow\bigl(\ket{3}-\ket{4}\bigr)/\sqrt{2}$. (Bottom) The gate corresponding to the right path was first applied to the condensate in numerical simulation and the lattice then held fixed. Subsequent propagation occurs along a single direction in accordance with $-4\hbar k_L$ momentum.}
    \label{fig:astheory}
\end{figure}

\begin{figure}[t]
    \centering
    \includegraphics[scale=0.5]{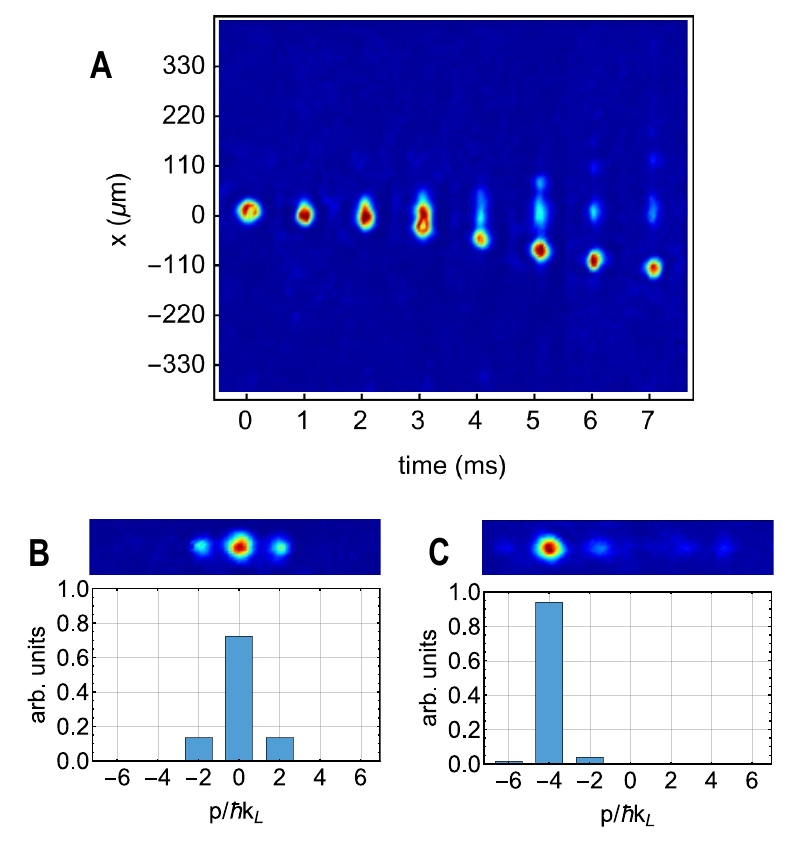}
    \caption{\textbf{Asymmetric beamsplitter  experiment:} {\bf (A)} and {\bf (B)}:~As in Fig.~\ref{fig:bsexpt} but applying the asymmetric beamsplitter shaking function in experiment and validating its operation. In {\bf (C)}, the bar graph for the momentum probability for $\bigl(\ket{3}-\ket{4}\bigr)/\sqrt{2}$ is in good agreement with that observed experimentally after the gate is applied.}
    \label{fig:asexpt}
    \vspace*{-1pc}
\end{figure}

The symmetric 50/50 beamsplitter can be generalized, since it is possible to design transformations that send unequal proportions of atoms along the two paths. The limiting case would be to use only one path, i.e., an asymmetric or 100/0 beamsplitter. In terms of Bloch state transitions, the target state for the asymmetric beamsplitter is \begin{equation}
\ket{0}\Rightarrow\frac1{\sqrt{2}}\bigl(\ket{3}+\ket{4}\bigr)
\end{equation}
to send all the atoms along one path, and
\begin{equation}
\ket{0}\Rightarrow\frac1{\sqrt{2}}\bigl(\ket{3}-\ket{4}\bigr)
\end{equation}
to send all the atoms along the other path. We derived a shaking function for this transformation, and in Fig.~\ref{fig:astheory} show the numerical simulation to confirm its functionality. In this simulation, the control function shakes the lattice for less than 200~$\mu$s, and then the lattice is held stationary to illustrate the subsequent evolution. Even though there are two separate design problems corresponding to the left and right paths, only one protocol has to be learned, since changing the sign $\phi(t)\Rightarrow-\phi(t)$ changes the direction.

As for the 50/50 beamsplitter, we also validated the operation of the asymmetric beamsplitter in experiment. The in situ evolution shown in Fig.~\ref{fig:asexpt} demonstrates that the atoms are primarily transferred into a~$-4\hbar k_L$ momentum component, and propagate through the lattice in a single direction for 7~ms. Small amounts of occupation are observed in other momentum orders, which is in agreement with the numerical simulation of Fig.~\ref{fig:astheory}. The anticipated gate behavior is confirmed by the momentum probabilities after~TOF.

\vspace*{-1pc}
\subsubsection*{\bf Mirror Gate}
\vspace*{-.75pc}

Enclosing a space-time area is necessary to build an inertial sensor and this requires a mirror so that the two distinct wave packet components that have been separated can be brought back together and recombined to produce the interference signal. The defining transformation is actually embedded in the design rules for the asymmetric gate just described, since $\bigl(\ket{3}+\ket{4}\bigr)/\sqrt{2}$ was the target state to travel in one direction, and $\bigl(\ket{3}-\ket{4}\bigr)/\sqrt{2}$ was the target state to travel in the other. Comparing these, and ignoring the global phase, it is evident that a mirror is well described as a device that satisfies one of two design options
\begin{equation}
\begin{array}{l}
\ket{3}\Rightarrow\ket{3}\\[.5pc]\ket{4}\Rightarrow-\ket{4}
\end{array}
\qquad\mbox{or}\qquad
\begin{array}{l}
\ket{3}\Rightarrow\ket{4}\\[.5pc]\ket{4}\Rightarrow-\ket{3}
\end{array}
\end{equation}
related by permuting the output labels. We refer to this as target operator design since for either option, two rules have to be simultaneously satisfied. This differentiates it from the previous two gates that were targeting a single state. Here we stay within a SU(2) subspace, i.e., the conduction band manifold $\{\ket{3},\ket{4}\}$, so this transformation can also be recognized as a single qubit gate. In fact, in the language of quantum information science, these possible transformations are referred to as the Pauli $Z$-gate and the Pauli $Y$-gate, respectively.
\begin{figure}[b]
    \centering
    \includegraphics[scale=0.4]{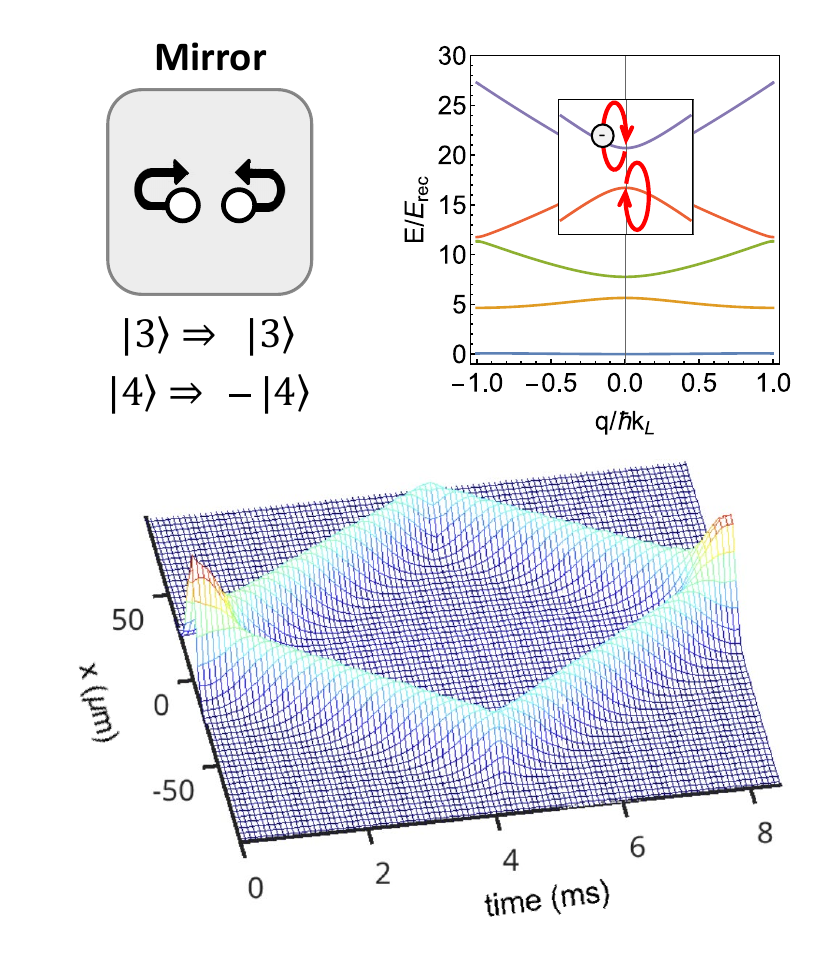}
    \vspace*{-1pc}
    \caption{\textbf{Mirror design:} (Top) A Bloch state transition for a mirror can be designed by negating the odd parity component of the wavefunction, $\ket{4}$, while leaving the even parity component, $\ket{3}$, unchanged. (Bottom) Numerical simulation for a sequence involving a beamsplitter, propagation, mirror, and further propagation.}
    \vspace*{-1pc}
    \label{fig:mirrorth}
\end{figure}
\begin{figure}[t]
    \centering
    \includegraphics[scale=0.5]{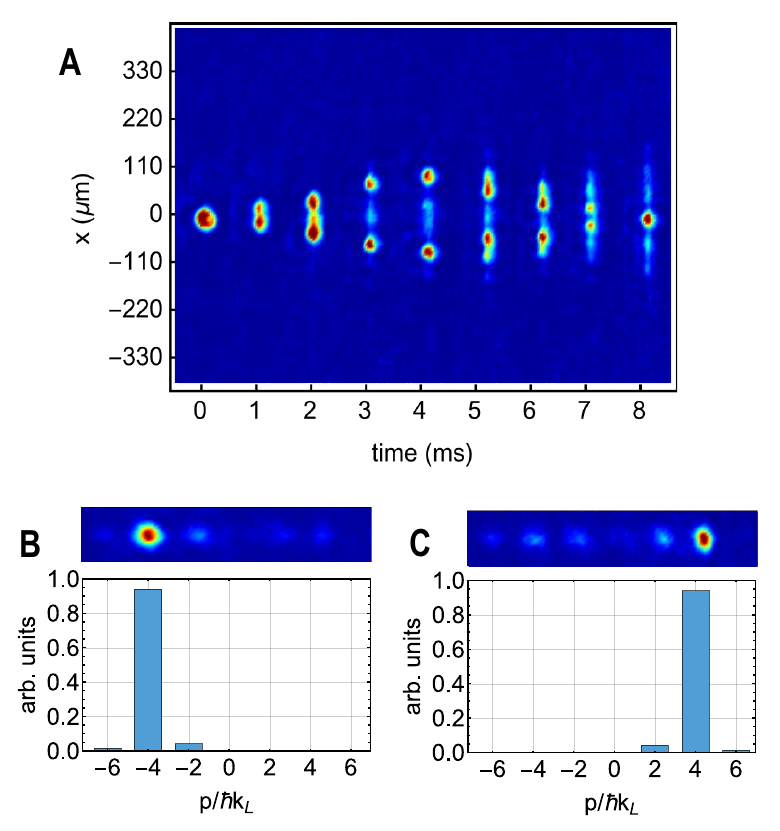}
    \caption{\textbf{Mirror experiment:} {\bf (A)} Experimental demonstration of the control function sequence that was simulated in Fig.~\ref{fig:mirrorth}, i.e., a beamsplitter, propagation, mirror applied at~4~ms, and propagation. In the TOF images, we used a different sequence in which the beamsplitter was replaced by an asymmetric beamsplitter to make the functionality clear. {\bf (B)} TOF image taken after an asymmetric beamsplitter and propagation but before the mirror protocol. {\bf (C)}~The same but after the mirror protocol.}
    \label{fig:mirrorexpt}
\end{figure}

In Fig.~\ref{fig:mirrorth}, the numerical simulation is shown using the shaking function that was found for this design. 
To make its functionality transparent, we solve the Schr\"odinger evolution for a composite waveform that involves stitching together a sequence of gates. We begin with a BEC of finite extent in the ground state of the lattice, apply the beamsplitter shaking function, evolve for 4 ms, apply the mirror protocol with the Bloch transformation described in this section, and then propagate for a further~4~ms. The reflection of the momentum components by the mirror is evident.

When we validated the operation in experiment (Fig.~\ref{fig:mirrorexpt}), the anticipated space-time diamond pattern is observed in the in situ image sequence, showing the wavepacket components separating along two paths, being reflected, and coming back together, all while being confined to the optical lattice. We point out that this is almost a complete Mach-Zehnder interferometer that misses only the final recombination step in which the interference is formed. In the experimental TOF validation of the momentum transformation, we replaced the symmetric beamsplitter with an asymmetric beamsplitter so that it was clear the atoms were being reflected. Similar results were seen for atoms sent along the other path.

\vspace*{-1pc}
\subsubsection*{\bf Conduction-Band Beamsplitter Gate}
\vspace*{-.75pc}

To make gradiometers that measure differential operators, we need another type of beamsplitter that symmetrically splits wavepacket components that are already propagating in the conduction band. Here the desired transformation can be defined by one of two alternate transformations; 
\begin{equation}
\begin{array}{l}
\ket{3}\Rightarrow\frac1{\sqrt{2}}\bigl(\ket{3}+\ket{4}\bigr)\\[1pc]\ket{4}\Rightarrow\frac1{\sqrt{2}}\bigl(\ket{3}-\ket{4}\bigr)
\end{array}\quad\mbox{or}\quad
\begin{array}{l}
\ket{3}\Rightarrow\frac1{\sqrt{2}}\bigl(\ket{3}-\ket{4}\bigr)\\[1pc]\ket{4}\Rightarrow\frac1{\sqrt{2}}\bigl(\ket{3}+\ket{4}\bigr)
\end{array}
\end{equation}
that operate entirely within the SU(2) manifold of conduction band states. These are also single qubit gates familiar in quantum information science where they are known as Hadamard $H$-gates.

\begin{figure}[t]
    \centering    \includegraphics[scale=0.4]{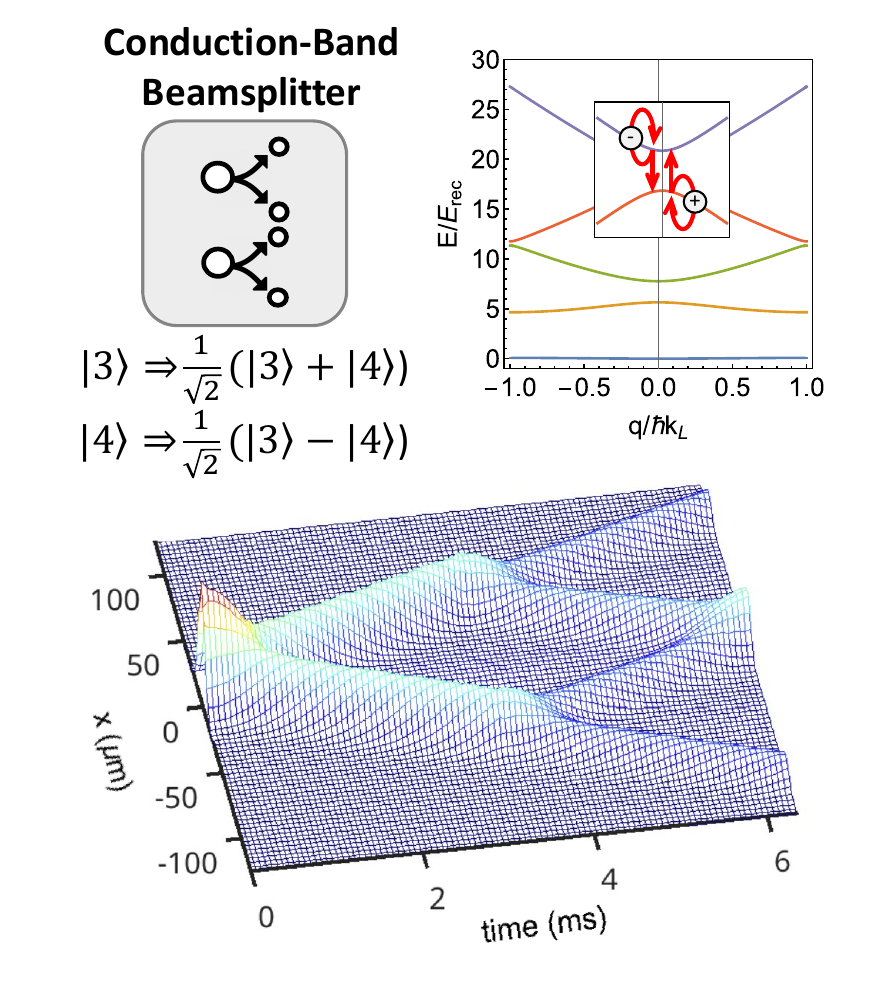}
    \vspace*{-1pc}
    \caption{\textbf{Conduction-band beamsplitter design:} (Top) The Bloch state transition for a conduction-band beamsplitter is the target unitary given by a Hadamard gate in the $\ket{3}$ and $\ket{4}$ subspace. (Bottom) Numerical simulation for a sequence of beamsplitter, propagation, conduction-band beamsplitter and further propagation.}
    \label{fig:cbbsth}
\end{figure}
\begin{figure}[ht!]
    \centering
    \includegraphics[scale=0.5]{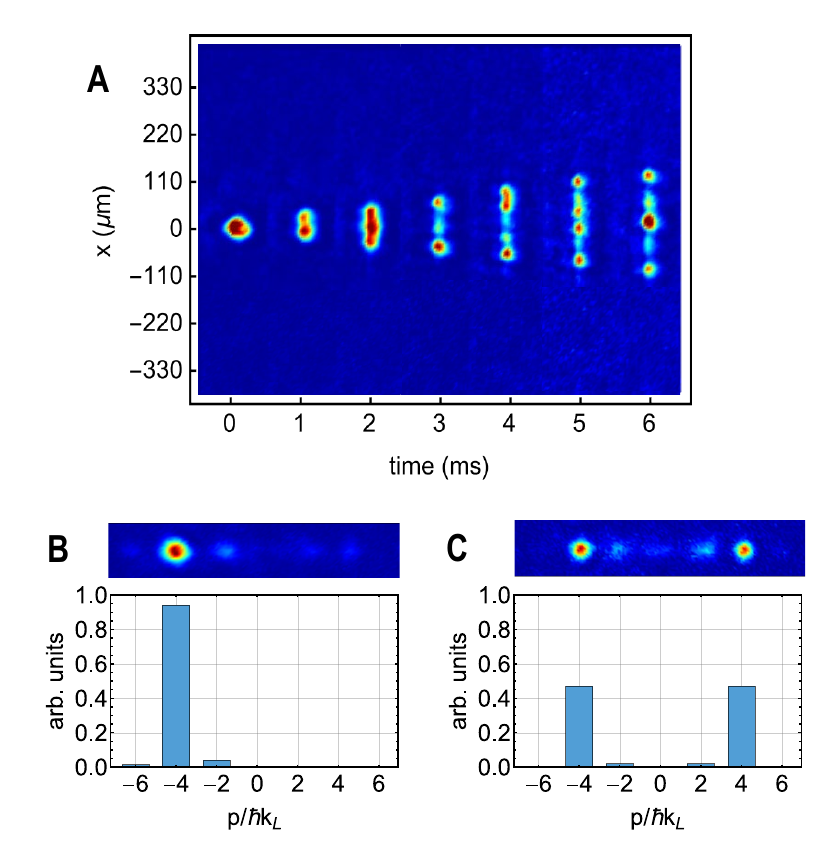}
    \caption{\textbf{Conduction-band beamsplitter experiment:} {\bf (A)}~Experimental demonstration of the sequence that was simulated in Fig.~\ref{fig:cbbsth}, i.e.,  beamsplitter and propagation, and the conduction-band beamsplitter applied at 3~ms with subsequent propagation. {\bf (B)} As for the mirror, here we replaced the initial symmetric beamsplitter with an asymmetric beamsplitter, and recorded the TOF image before the conduction-band beamsplitter was applied. {\bf (C)}~The same, but after the conduction-band beamsplitter. The state that is principally $-4\hbar k_L$ is divided into $\pm4\hbar k_L$ by the gate.}
    \label{fig:cbbsexp}
\end{figure}

We generated a shaking function for this target operator and demonstrated its application by numerical simulation, as shown in Fig.~\ref{fig:cbbsth}. In this simulation we solved the Schr\"odinger evolution for a sequence involving a beamsplitter, propagation for 3~ms, conduction-band beamsplitter, and additional propagation for a further~3~ms. The splitting of wave packets already propagating in the conduction band is evident. 

The corresponding experimental demonstration for this sequence is given in Fig.~\ref{fig:cbbsexp} showing that the sequence functions as anticipated. The initial beamsplitter divides the BEC into two parts, and the atom clouds are split a second time by the application of the conduction-band beamsplitter at 3~ms. At the end of the sequence, three peaks are visible in the in situ image, with a large central peak that is due to atoms that have taken either the upper or lower middle paths meeting in the center. When observing the experimental TOF image, as for the mirror, we employed an asymmetric beamsplitter to start with so that the splitting in the conduction band was evident in the resulting momentum probabilities.

\vspace*{-1pc}
\subsubsection*{\bf Split \& Hold Gate}
\vspace*{-.75pc}

\begin{figure}[b]
    \centering
    \includegraphics[scale=0.4]{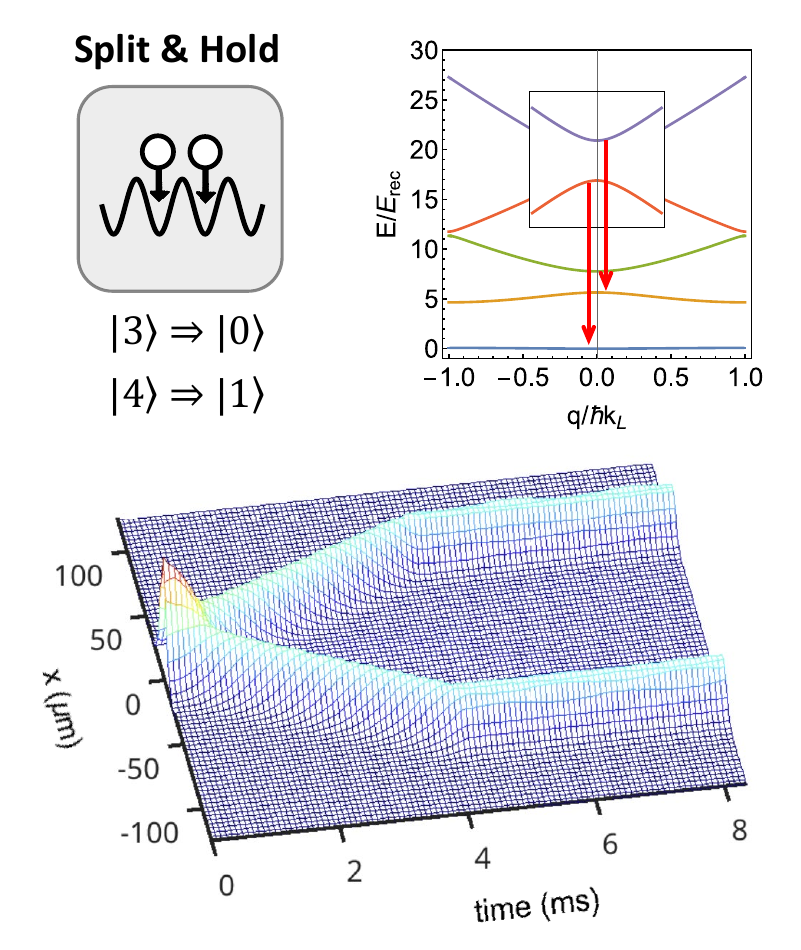}
    \vspace*{-1pc}
    \caption{\textbf{Split and Hold design:} (Top) The Bloch state transition for split \& hold is a target unitary that takes the atoms from the two conduction band states and maps them to the lowest two valence band states. (Bottom) To see the functionality clearly, the simulation applies the sequence of beamsplitter, propagation, and then split \& hold at 4~ms, with subsequent evolution showing the atoms no longer moving between sites.}
    \label{fig:shth}
\end{figure}

\begin{figure}[htb]
    \centering
    \includegraphics[scale=0.5]{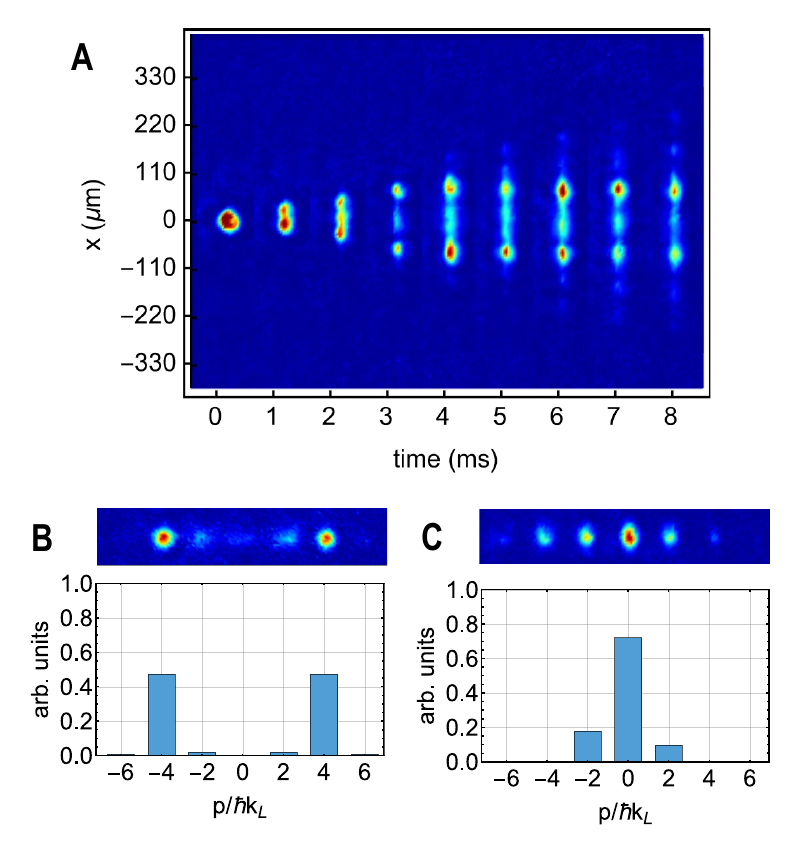}
    \caption{\textbf{Split \& hold experiment:} {\bf (A)}~Experimental demonstration of the sequence in Fig.~\ref{fig:shth}, i.e., beamsplitter, propagation, split \& hold at~4~ms, and further propagation for~4~ms. {\bf (B)} TOF image for the momentum probability before the split \& hold gate. {\bf (C)}~The same but after the split \& hold gate. Although only one image is shown, the image is not stationary. Instead the $\pm2\hbar k_L$ probabilities periodically exchange population at the vibrational frequency as a function of time after the gate is applied.}
    \label{fig:shexp}
\end{figure}

A possibility that is created by leaving the optical lattice potential on during the entire interferometry sequence is the ability to transfer atoms back into the non-propagating valence band. A split \& hold gate is a target operator transformation that performs this by the following two alternate design options
\begin{equation}
\begin{array}{l}
\ket{3}\Rightarrow\ket{0}\\[.5pc]\ket{4}\Rightarrow\ket{1}
\end{array}\qquad\mbox{or}\qquad
\begin{array}{l}
\ket{3}\Rightarrow\ket{1}\\[.5pc]\ket{4}\Rightarrow\ket{0}
\end{array}
\end{equation}
representing a conduction to valence band mapping. This offers the intriguing possibility of transforming atoms during an interferometry sequence into states where they can be held and accumulate inertial phase for extended periods without requiring an increase in the footprint of the device. Confining atoms in an optical lattice at controllable distances is comparable to the cavity-mediated interferometers~\cite{PhysRevLett.114.100405, Panda2024}, which offer enhanced interrogation time and increased immunity to high frequency phase noise. 

Once confined in the valence states of the lattice after the application of the split \& hold gate, the wavefunction is no longer in an eigenstate. Instead the amplitudes for $\ket{0}$ and $\ket{1}$ will evolve with a relative phase that accumulates at a rate dependent on the energy spacing of the lowest two levels. This can equivalently be viewed by the manifestation that the total probability density will oscillate back and forward in the potential wells. Since the optical lattice potential is sinusoidal and the bottom of the potential wells is approximately harmonic, this is often referred to as a vibrational oscillation in analogy with molecules.

The numerical simulation for the shaking function that was found for this Bloch transformation is shown in Fig.~\ref{fig:shth}. The sequence used to demonstrate the functionality of the split \& hold protocol is the application of a beamsplitter, propagation for 4~ms, split \& hold, and then an additional 4~ms of hold time to view any subsequent evolution. Before the split \& hold gate, the atoms transport through the lattice in the conduction band, and after the split \& hold they are localized in the potential wells in the valence band states where they are frozen in place and no longer travel through the lattice.

Experimental results of the application of the split \& hold gate is shown in Fig.~\ref{fig:shexp}. After a beamsplitter and subsequent 4 ms propagation, the split \& hold gate is applied, and the atoms are held in place for a further~4~ms hold. Atoms that are not faithfully returned to the trapped bands are faintly discernible by their increasing separation at long times. 

\vspace*{-1pc}
\subsubsection*{\bf Echo Gate}
\vspace*{-.75pc}

\begin{figure}[t]
    \centering
    \includegraphics[scale=0.4]{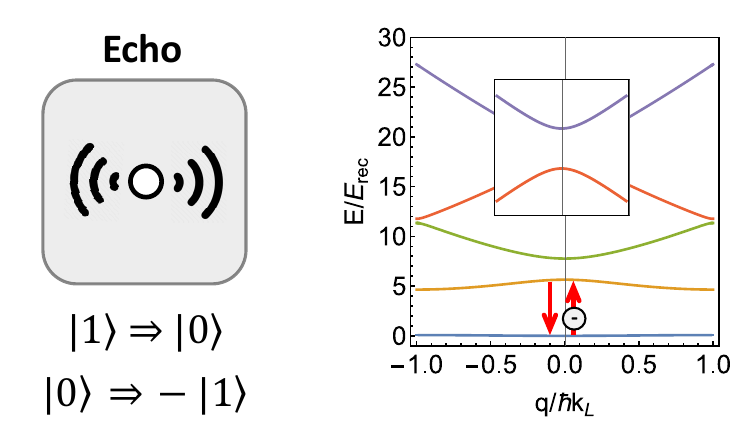}
    \caption{\textbf{Echo design:} The echo gate is simply a mirror in the valence band subspace of $\{\ket{0},\ket{1}\}$.} 
    \label{fig:echoth}
\end{figure}
\begin{figure}[b]
    \centering
    \includegraphics[scale=0.5]{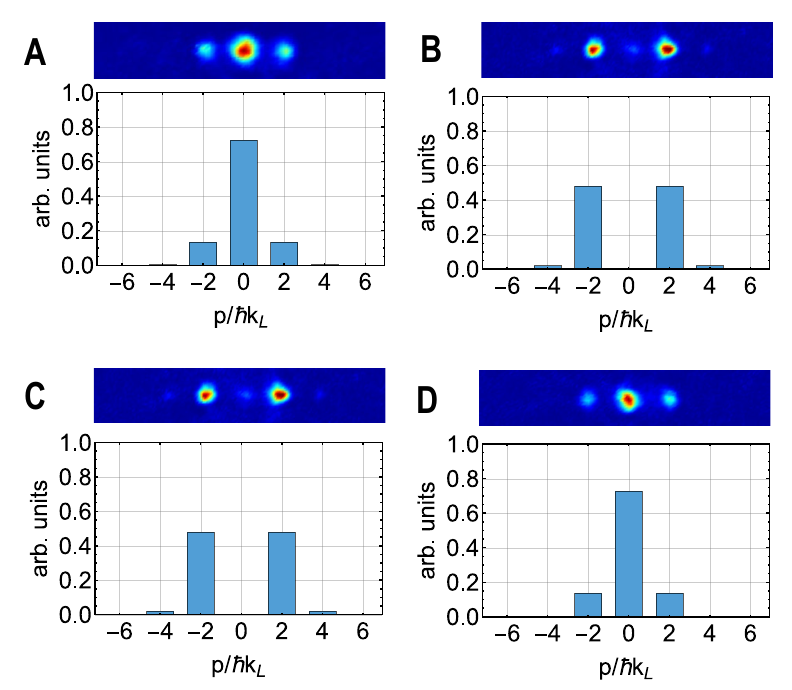}
    \caption{\textbf{Echo experiment:} We do not provide in situ images for this case since the atoms are trapped in the potential wells and do not move. Instead, we provide an experimental demonstration of its function by showing a sequence of TOF images of the momentum components. We begin with a condensate in the ground state $\ket{0}$, shown in {\bf (A)}, then apply the echo protocol, giving the result shown in {\bf (B)}, to produce the first vibrational state $\ket{1}$. Then we start in the first vibrational state, shown in {\bf (C)}, apply the echo protocol again, and return the atoms to the ground state, shown in {\bf (D)}.}
    \label{fig:echoexp}
\end{figure}

The echo gate allows one to confine atoms in lattice potential wells with the split \& hold protocol and then release them in a manner that does not require a specific hold time. If the exact vibrational spacing of the lowest two valence states were perfectly known, and the hold time could be adjusted to a half-integer of the vibrational period, then the echo gate would not be needed. However, this may not be possible even in principle since spatial variations of the vibrational spacing due to lattice depth variation or other effects may lead to inhomogeneous dephasing. In this case, the necessary hold time designed into the protocol may not be single valued for different lattice sites. 

To provide an effective solution to this problem, we draw inspiration from spin-echo protocols~\cite{PhysRevLett.82.2417, LEVITT198165} that introduce a $\pi$-pulse at the middle of the interferometer sequence to compensate for inhomogeneous dephasing. This design protocol is defined by the two possibilities;
\begin{equation}
\begin{array}{l}
\ket{0}\Rightarrow\ket{0}\\[.5pc]\ket{1}\Rightarrow-\ket{1}
\end{array}\qquad\mbox{or}\qquad
\begin{array}{l}
\ket{0}\Rightarrow\ket{1}\\[.5pc]\ket{1}\Rightarrow-\ket{0}
\end{array}
\end{equation}
which is simply a mirror in the valence band which reflects the oscillation of the wavepackets in the potential wells. In the terminology of quantum information science, the echo gate can be recognized as a single qubit Pauli $Z$-gate or Pauli $Y$-gate in the $\ket{0}$ and $\ket{1}$ subspace. The intended application of the echo gate is to apply it at the half-time of a hold phase which we will see later through its application in various sensing circuits. 

We have solved for a shaking function for the echo transformation, validated it in numerical simulation, and demonstrated its function in experiment as shown in Fig.~\ref{fig:echoexp}. In this sequence, an initial BEC was transformed from the ground state into the first vibrational state by the application of an echo gate, and then brought back to the ground state by a second application.

\section{Metrological Universality}
\vspace*{-.75pc}
Thus far, we have introduced the necessary individual gates for a programmable matter-wave interferometer. In subsequent sections, we will describe how to sequence these gates and make protocols that can detect a wide variety of signals. However, first we introduce the important notion of a universal gate set in the context of quantum metrology---a concept inspired by the architecture of quantum computation.

In quantum computing, if a finite set of gates can carry out any unitary operation to an arbitrarily high fidelity, then this set is said to be a ``universal gate set''. 
This is significant since any quantum algorithm can be performed by a universal gate set. This inspires our proposed definition of a metrologically universal gate set in the context of a programmable inertial sensor. We define metrological universality here as the ability to detect any signal in the form of a force and any local linear map of that force. Vector differentials such as curls, divergences, and gradients are all local linear maps, and so this notion unifies an entire class of inertial sensing that includes vector accelerometry and gravimetry, multi-axis rotation sensing with gyroscopes, tensor gradiometry, magnetometry and electric field measurements, among others.

A gate set being metrologically universal is a weaker condition than it being (computationally) universal.
In quantum computing, one uses gates to build arbitrary input-output {\em quantum operations}. 
In quantum metrology, one uses gates to construct input-output protocols that embed sensitivity to {\em arbitrary signals}. 
However, despite the differences in theoretical construct, metrological and computational universality are functionally similar. They both involve an operational framework in which one sequences together elementary unitary gates to build circuits that can be used for universal functionality.

In the following sections, we demonstrate the universality of our gate set through example, that is, by the explicit construction of the important sensing circuits to measure accelerations, gravity gradients, and rotations. We calculate each protocol's response to an applied signal through the path integral formalism of matter-wave interferometry~\cite{storey1994feynman}, and numerically show that these phases are detectable by our momentum observables. 

\vspace*{-1pc}
\section{Sensing Sequences}
\vspace*{-.75pc}

We explicitly detail two classes of accelerometers, two classes of gravity gradiometers, and a matterwave gyroscope, each comprised of a sequence of elementary matterwave gates. For each sensing sequence, we present the time-ordered circuit and numerically simulate the space-time evolution of the atomic wavepacket accounting for the full atom-lattice dynamics. The response of these protocols as a function of the parameter of interest is shown by the momentum-space interference pattern that results. In this interference pattern, the probability amplitudes for the discrete momentum orders can be considered to be the output channels of a multiport sensor from which the inertial phase is extracted.

For each sensor, we will provide an anticipated device sensitivity and its dependence on parameters, which can be calculated analytically from the path integral. For example, for a two-path atom interferometer, the relative inertial phase $\Delta\phi$ that is accumulated between the two arms is given by:
\begin{equation}
\Delta \phi = \frac{1}{\hbar} ( S_+ - S_- )
\end{equation}
written in terms of the action
\begin{equation}
S_\pm= \int_0^\tau \Bigl\{ L_0(x_\pm, {\dot x}_\pm;t) - V_I(x_\pm, {\dot x}_\pm;t) \Bigr\}\, dt
\label{eq:PathInt}
\end{equation}
where $\pm$ labels the classical paths.
These paths are integrated over time~$t$ of the sequence, up to the terminal time~$\tau$. At each point in time, the action depends on the free Lagrangian of the matterwave-lattice system $L_0$ and interaction potential that we wish to sense $V_I$. To find analytical expressions for the phases, we integrate the classical paths as given by the stationary phase approximation~\cite{storey1994feynman}. We do this in limit of long propagation times where the gates are effectively instantaneous.

From the numerical simulations for the complete sensor sequence, we calculate the resultant momentum probability fringes for each possible momentum diffraction order~$m$, included as insets to the sensitivity scaling plots. The sensitivity scaling of each protocol is found by first calculating the classical Fisher information~(CFI)~\cite{Barndorff-Nielsen_2000, PhysRevA.62.012107}, which characterizes the information gained about the value of a parameter $\beta$ from a single measurement $m$. The CFI is defined as 
\begin{eqnarray}
    {\cal I}(\beta) &=& \mathbb{E}_{\beta}\left[\bigg(\frac{\partial}{\partial\beta}\log P(m|\beta)\bigg)^2\right]\nonumber\\
    &=& \sum_m \frac{1}{P(m|\beta)}\left[\frac{\partial P(m|\beta)}{\partial \beta}\right]^2
    \label{eq:2::fisher}
\end{eqnarray}
where $\mathbb{E}_\beta$ is the expectation value evaluated at $\beta$,  and $P(m|\beta)$ is the conditional probability of getting order~$m$ given $\beta$. This expression is for a single atom wavefunction; the sensitivity for an ensemble is given by 
\begin{equation}
   \Delta\beta=\frac1{\sqrt{N{\cal I}(\beta)}}
\end{equation}
where $N$ is the number of independent trials that can be extracted from a single absorption image. In this paper we will always assume $N=1000$, a typical value measured in the current experiment~\cite{ledesma2024vector}.
Since the CFI is not constant across the entire scan, we plot the obtainable sensitivity as a shaded region, bounded by the maximum and minimum over the calculated scan range. The sensitivity improves as the square root of the shot number (number of experiments) due to statistical averaging.

\vspace*{-1pc}
\subsubsection*{\bf Accelerometer}
\vspace*{-.75pc}

\begin{figure}[t]
    \centering
    \includegraphics[scale=0.45]{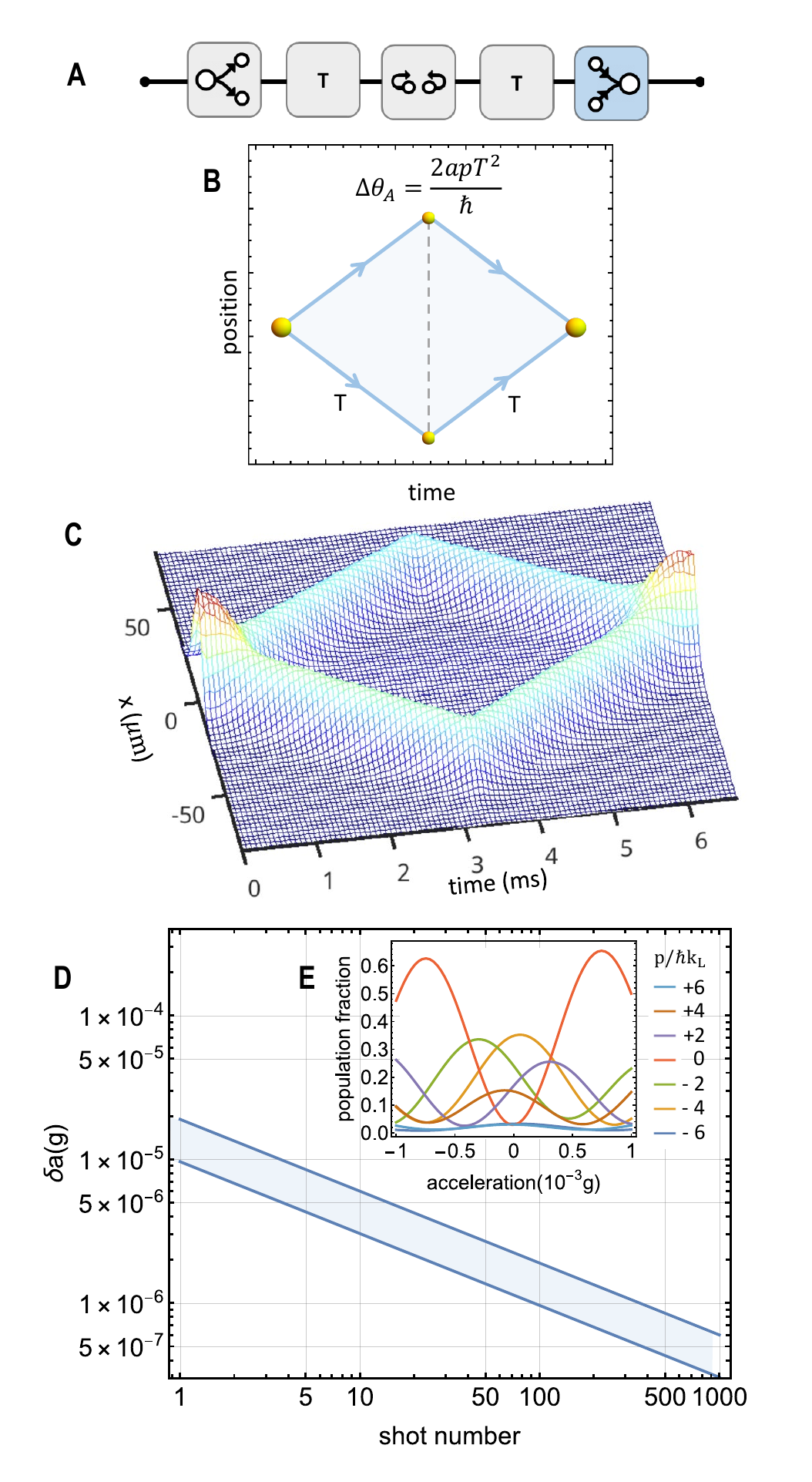} \caption{\textbf{Accelerometer:} \textbf{(A)} Matterwave circuit for an accelerometer with each tile representing a gate. \textbf{(B)} Space-time plan view of the wavepacket geometry. \textbf{(C)}~Numerical simulation of the complete accelerometer protocol, where the control function was sequenced by stitching together the shaking solutions for the various gates. The propagation time was $T= 3$~ms. \textbf{(D)} Short-term sensitivity, with shot number representing repeated experiments. The shaded region bridges the minimum and maximum sensitivity over the scan range. \textbf{(E)} Image of the 7 output momentum state fringes that are obtained by applying an acceleration scan range of $-0.001g$ to $+0.001g$, where~$g$ is the average gravitational acceleration at the Earth's surface. }
    \label{fig: Accelerometer}
    \vskip-1pc
\end{figure}

The accelerometer sequence is a beamsplitter, mirror, and recombiner with variable transport delays between operations, as shown in the matterwave circuit seen in Fig.~\ref{fig: Accelerometer}A. A simplified plan view of the 2D space-time area enclosed by the atoms during the applied sequence can be seen in Fig. \ref{fig: Accelerometer}B. The accelerometer control function was created by stitching together the shaking sequences that were learned for the various components, where for the recombiner we used the reverse beamsplitter, and included delays of $T=3$~ms between gates during which $\phi(t)=0$. Starting from a BEC in the ground state of the lattice and numerically simulating this control function results in the wavefunction evolution shown in  Fig.~\ref{fig: Accelerometer}C, where the familiar diamond pattern of the matterwave interferometer can clearly be seen. 

For an acceleration $a$, the potential that we wish to sense is $V_I(x) = m a x$, and the relative phase accumulated between the upper and lower arms of the accelerometer according to Eq.~\ref{eq:PathInt} is then given by
\begin{equation}
\Delta \theta_{A} = \frac{2 a p T^2}{\hbar}
\label{eq:accelformula}
\end{equation}
where $p = 4 \hbar k_L$. This formula has an intuitive interpretation as the product of the acceleration $a$ with the space-time area of the diamond in units of~$\hbar/m$. It arises from the fact that the components of the wavepacket in the two arms of the accelerometer travel through two different parts of the potential field which manifests as a phase difference in the matterwave amplitudes at recombination. We point out that this formula is fundamentally no different from a conventional light-pulse accelerometer. This design formula is a direct consequence of the matterwave physics~\cite{CLAUSER1988262} and de Broglie~\cite{de1929wave} relations that underpin all of the sensing devices.

We scan the output as a function of $a$ by adding $V_I(x)$ to the atom-lattice Hamiltonian and numerically simulating the evolution for the accelerometer control function. This allows us to calculate the expected sensitivity from the resulting classical Fisher information. We then plot the sensitivity as a function of repeated experiments (shot number), where we show the result in Fig.~\ref{fig: Accelerometer}D. The raw fringes for each momentum component used to calculate the CFI are also provided for reference and are given in the inset Fig.~\ref{fig: Accelerometer}E. The observed fringe period is in good agreement with Eq.~(\ref{eq:accelformula}).

\vspace*{-1pc}
\subsubsection*{\bf Accelerometer with Hold}
\vspace*{-.75pc}

\begin{figure}[t]
    \centering
    \includegraphics[scale=0.45]{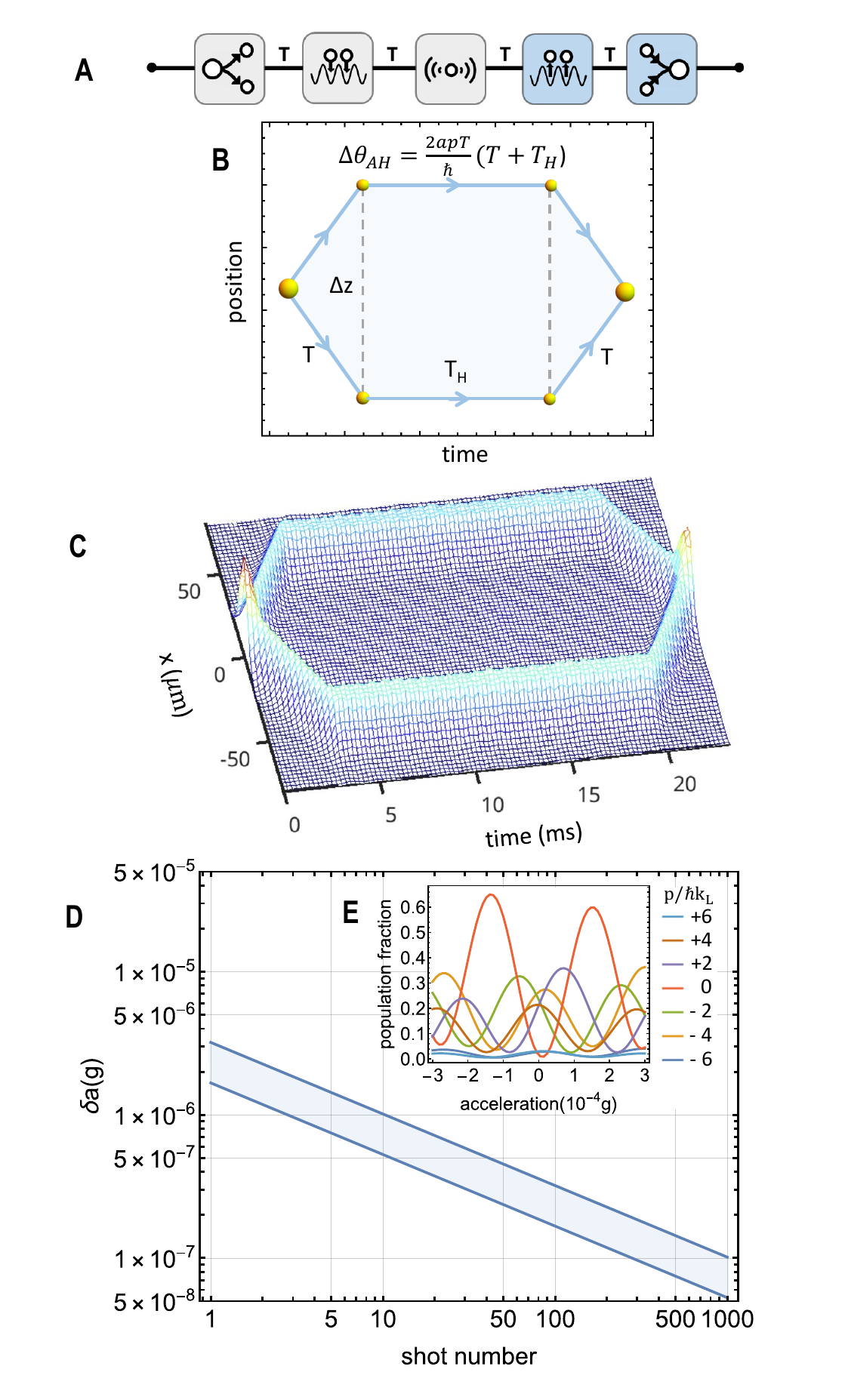}    \caption{\textbf{Accelerometer with Hold:} \textbf{(A)} Circuit representation of an accelerometer with hold protocol. The gates are beamsplitter, split \& hold, and echo, with blue indicating reverse operation. \textbf{(B)} Plan view of the wavepacket evolution. \textbf{(C)} As in Fig.~\ref{fig: Accelerometer}C, but for the accelerometer with hold circuit and using a hold time of 16~ms total. \textbf{(D)} Plot of the short-term sensitivity that results, the shaded region indicating the maximum and minimum CFI over the scan range. \textbf{(E)} Scan of the 7 momentum state fringes obtained for $a$ in the range $-0.0003g$ to $0.0003$g.}
    \label{fig: Accelerometer with Hold}
\end{figure}

An alternative approach to enhance the sensitivity to applied forces is a hybrid accelerometer that initially separates the atomic wavepacket and but then holds the atoms at a fixed distance apart for extended interrogation times. This decouples the sensitivity from the physical dimensions of the sensor because the space-time area can be increased by simply holding the atoms for longer times instead of having them separate further apart. Since the gates are sequenced, this hold time can be programmed in software. The matterwave circuit and its associated plan view are given in Fig.~\ref{fig: Accelerometer with Hold}A and~B. The sequence is a beamsplitter, split \& hold, echo, reverse split \& hold, and recombiner, with transport times occurring while atoms are in the conduction band, and hold times while they are in the valence band.

The function of the echo pulse should now be transparent. The $\ket{1}$ state is of higher energy than the $\ket{0}$ state, and so during a hold, the relative phase of wavepacket amplitudes in these states change in time. The echo pulse placed at the midpoint means that during the second part of the hold this phase accumulation is unwound. The result is that the combination of split \& hold, echo, reverse split \& hold replaces the mirror in the conventional geometry. 

In the accelerometer with hold, $V_I(x)$ is the same as in the normal accelerometer, and the relative phase accumulated between the upper and lower arms is
\begin{equation}
\Delta \theta_{AH} = \frac{2 a p T (T+T_H)}{\hbar}
\label{eq:accelhold}
\end{equation}
or equivalently $\Delta \theta_{AH} = \Delta \theta_{A} + \bigl(2 a p T T_H\bigr)/\hbar$, where the additional term represents the increased space-time area that increases linearly with the hold time.

The numerical simulation of the accelerometer with hold protocol that stitches together the gate shaking sequences for a transport time of~3~ms and a hold time of 16~ms is shown in Fig.~\ref{fig: Accelerometer with Hold}C, and the resulting scan is shown in Fig.~\ref{fig: Accelerometer with Hold}D and E. This can be compared with the conventional accelerometer geometry in Fig.~\ref{fig: Accelerometer}D and E. From the variation of the momentum fringes, it is apparent that the device has increased sensitivity and the results agree well with Eq.~(\ref{eq:accelhold}).

\vspace*{-1pc}
\subsubsection*{\bf Gradiometer}
\vspace*{-.75pc}

Gradiometers are sensitive to the gradient of an inertial force. This measurement is typically accomplished by comparing accelerometers or gravimeters that are some distance apart. In traditional atom gradiometers, two spatially separated atomic samples are simultaneously split, mirrored, and recombined, using common light pulses, such that readout of the interference fringes provides information about the gravity gradient. In contrast, what we demonstrate here is a gradiometer that self-references, i.e., enabling the measurement of gradients from a single atomic source. 

\begin{figure}[b]
    \centering
    \includegraphics[scale=0.45]{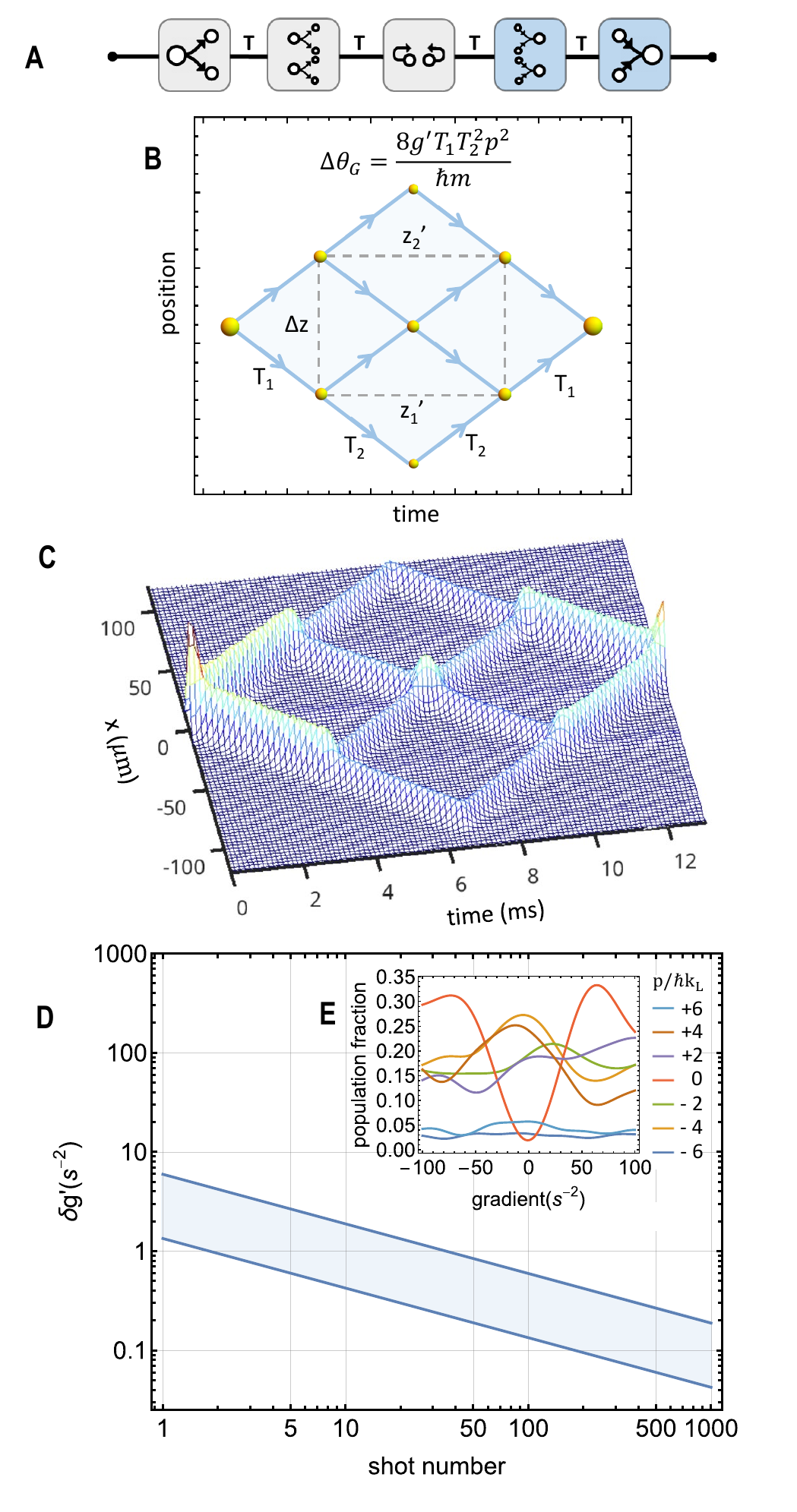}    \caption{\textbf{Gradiometer:} \textbf{(A)} Circuit representation of a gradiometer protocol. The symbols are beamsplitter, conduction-band beamsplitter and mirror, with blue indicating reverse operation. \textbf{(B)} Plan view of the wavepacket evolution. \textbf{(C)} Numerical simulation using the gradiometer protocol that is made by stitching together the shaking functions for each gate, and with a transport time in all stages of $T=3$ms. \textbf{(D)}~Plot of the short-term sensitivity, the shaded region indicating the CFI spread over the scan range. \textbf{(E)} Image of the 7 momentum fringes that are obtained from applying a gradient scan. }
    \label{fig: Gravity Gradiometer}
\end{figure}

The necessary beamsplitter, conduction-band beamsplitter, mirror, reverse conduction-band beamsplitter, and reverse beamsplitter gates are sequenced in the sensing circuit of Fig.~\ref{fig: Gravity Gradiometer}A. Following a beamsplitter and propagation, application of the conduction-band beamsplitter splits the wavepacket into four distinct paths that propagate further. Application of a mirror, reverse conduction-band beamsplitter, and reverse beamsplitter manipulates the wavepackets to enclose two spatially separated space-time diamonds, and then recombines all the components. This can be seen in the plan view given in Fig.~\ref{fig: Gravity Gradiometer}B. The recombination creates an interference pattern that reflects the relative accumulated phase difference between the upper and lower interferometers, thus implementing a self-referenced gradient measurement. 

In the gradiometer, the potential that we sense is $V_I(x) = m g' x^2$, where $g'$ is found from the gradient of the acceleration field, and $x^2$ corresponds to the local behavior of this first derivative.
The relative phase accumulated between the upper and lower diamonds of the gradiometer and measured at the recombination stage is
found by computing the action along four paths;
\begin{equation}
\Delta\theta_G = \frac1{\hbar}\bigl(S_{\rm uu} - S_{\rm ul} - S_{\rm lu} + S_{\rm ll}\bigr)
\end{equation}
with $u\equiv\mbox{upper}$ and $l\equiv\mbox{lower}$, the first index the result of the beamsplitter, and the second index the result of the conduction-band beamsplitter. Evaluating this gives
\begin{equation}
\Delta \theta_{G} = \frac{8 g' T_1 T_2^2 p^2 }{\hbar m}.
\label{eq:gradphase}
\end{equation}
where $T_1$ is the propagation time between the beamsplitter and conduction-band beamsplitter, and $T_2$ is the propagation time between the conduction-band beamsplitter and the mirror. The greatest sensitivity for fixed total time is achieved for $T_2=2T_1$ indicating that it is better to have the diamonds overlap than to be separated in space as is typical in a conventional gravity gradiometer arrangement. Another important result is that the sensitivity increases rapidly in the gradiometer as the third power of the time.

The numerically simulated wavepacket evolution using the gradiometer protocol that is created by stitching together the shaking waveforms represented by the matterwave circuit seen in Fig.~\ref{fig: Gravity Gradiometer}A is given in Fig.~\ref{fig: Gravity Gradiometer}C for transport times of 3~ms in all stages. The scaling sensitivity and momentum state fringes are given in Fig.~\ref{fig: Gravity Gradiometer}D and E. The variation of the fringes agrees well and can be predicted by Eq.~(\ref{eq:gradphase}). 

Finally we point out that although we are not providing simulations of the multidimensional versions of these sensors in this paper for the case of brevity, the second beamsplitter (conduction-band beamsplitter) can obviously operate in another dimension to the original beamsplitter. This would enable measurements of the off-diagonal terms in the gradient tensor.

\subsubsection*{\bf Gradiometer with Hold}
\vspace*{-.75pc}

\begin{figure}[b]
    \centering
    \includegraphics[scale=0.40]{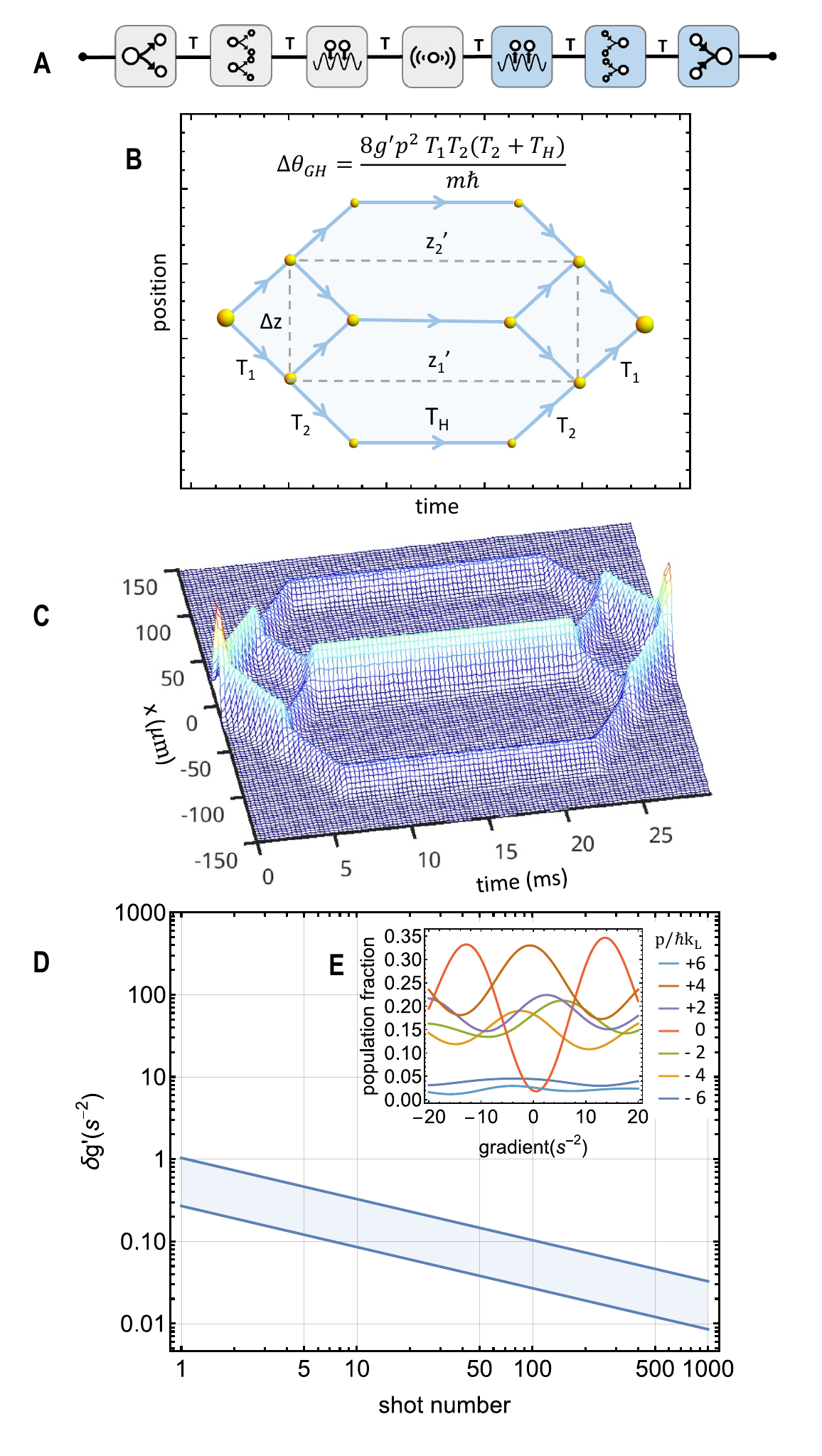} \caption{\textbf{Gradiometer with hold:} \textbf{(A)} Circuit representation of a gradiometer with hold sequence. The symbols are beamsplitter, conduction-band beamsplitter, split \& hold, and echo, with blue indicating reverse operation. \textbf{(B)} Plan view of the wavepacket evolution. \textbf{(C)} Numerical simulation of the total waveform created by stitching together the shaking solutions for the component gates. The transport time was 3~ms and the hold time was 8~ms. \textbf{(D)} Plot of short-term sensitivity. The shaded region represents the variation of CFI over the scan range. \textbf{(E)} Image of the 7 momentum state fringes that are obtained from applying a gradient scan.}
    \label{fig: Gravity Gradiometer with Hold}
\end{figure}

Performing gradiometry with a designed hold stage offers the same benefits as the accelerometer with hold by allowing us to extend the interrogation time without needing to increase the size of the device. As we did for the accelerometer, the implementation is straightforward and simply involves replacing the mirror with a split \& hold component set. The total sequence for a gradiometer with hold is beamsplitter, conduction-band beamspitter, split \& hold, echo, reverse split \& hold, reverse conduction-band beamsplitter, and recombiner, all intermediated by transport and hold steps. This sequence along with the plan view is shown in Fig.~\ref{fig: Gravity Gradiometer with Hold}A and B.

In the gravity gradiometer with a hold, the potential that we sense is again the same as the normal gradiometer.
The relative phase accumulated between the upper and lower diamonds of the gradiometer is
\begin{equation}
\Delta \theta_{GH} = \frac{8 g' T_1 T_2 (T_2 + T_H) p^2 }{\hbar m}.
\label{eq:gradhold}
\end{equation}
or $\Delta \theta_{GH} = \Delta \theta_G + \bigl(8 g' T_1 T_2 T_H p^2\bigr)/\hbar m$. Similar to the accelerometer with hold, the added term arises from increasing the space-time area by storing the atoms in place. 

\begin{figure*}[t]
    \centering
    \includegraphics[scale=.44]{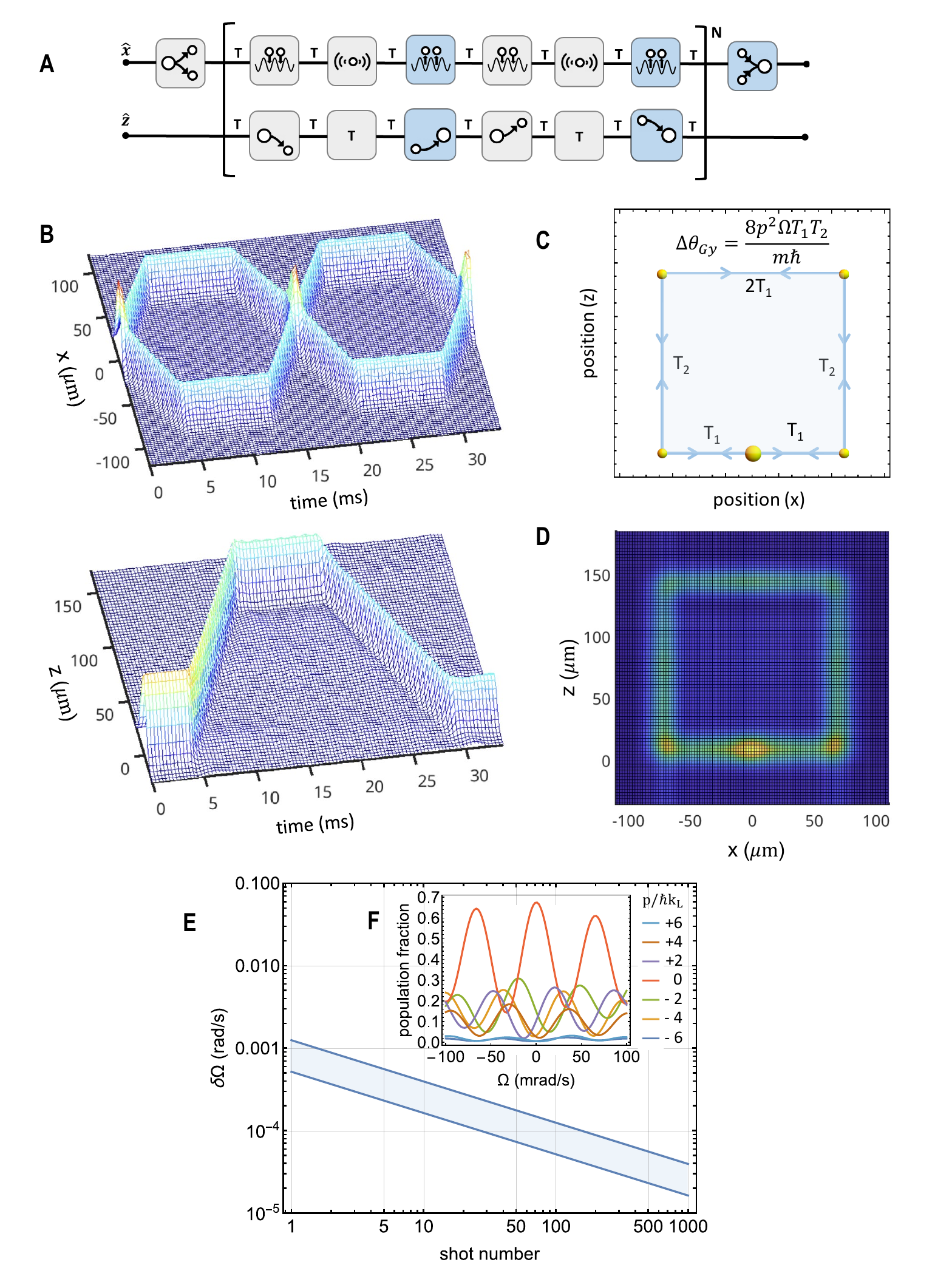}
    \caption{\textbf{Gyroscope:} \textbf{(A)} Circuit representation of a gyroscope sequence, where {\bf N} is the number of loops that the atoms undergo. \textbf{(B)} Space-time diagram of the wavepacket evolution over the course of the sequence in both the $x$ and $z$ dimensions. Each of the control functions for the gates in each dimension were stitched together and the two-dimensional Schr\"odinger equation was numerically solved. \textbf{(C)} Plan view of the gyroscope indicating the time symbols used. \textbf{(D)} Probability density averaged over the full time of the gyroscope sensor sequence. (\textbf{E}) Short-term sensitivity. The shaded region represents the variation of CFI over the scan range. \textbf{(F)} Image of the 7 momentum state fringes that are obtained from applying a rotation scan.}
    \label{fig:Gyroscope}
\end{figure*}

After combining the component shaking functions by stitching them together in the circuit sequence to form a gradiometer with hold control function, we solve numerically the Schr\"odinger equation and the result is given in Fig.~\ref{fig: Gravity Gradiometer with Hold}C. The simulation was performed with 3~ms transport time in all stages and 8~ms hold time. The associated sensitivity scan is in Fig.~\ref{fig: Gravity Gradiometer with Hold}D and E. Note the increased sensitivity with respect to the gradiometer without hold (Fig.~\ref{fig: Gravity Gradiometer}C). The observed momentum fringes agree well with Eq.~(\ref{eq:gradhold}).

\vspace*{-1pc}
\subsubsection*{\bf Gyroscope}
\vspace*{-.75pc}

Creating an atomic gyroscope sequence requires two components of a wave packet to traverse the perimeter of an area in opposite directions. This creates the same topology of atomic paths as seen in, for example, the Aharonov-Bohm effect, although the signal is generated here by rotation rather than the vector potential associated with a magnetic field. There are several simple configurations of gyroscopes that we can implement with our universal gate set. We will show one example here that is in the spirit of this paper by using only atom optic components.

Our atomic gyroscope sequence is shown in Fig.~\ref{fig:Gyroscope}A. This is fundamentally a two-dimensional sensor that therefore requires two lines in the matterwave circuit, one for each dimension $x$ and $z$. It involves sequenced gates with beamsplitters, split \& holds, echos, and asymmetric beamsplitters, along with their inverses. The sequence within the brackets represents a single loop, which may be successively applied ``{\bf N}" times to increase the total enclosed area of the sensor~\cite{Schubert2021}. The atoms are split in $x$, held in $x$ while being transported in $z$, held in $z$ while being released in $x$ to exchange sides, again held in $x$ while being transported back in $z$, and finally held in $z$ while being released in $x$ so the components can be recombined to produce interference. The numerical simulation of the action of all the control functions for the gates is given in Fig.~\ref{fig:Gyroscope}B with a 3~ms half-side propagation time. The total enclosed area is 0.010~mm$^2$ per loop. The 2D plan view is given in Fig.~\ref{fig:Gyroscope}C showing the geometry, where in this case we show a two-dimensional space diagram and label the times it takes atoms to traverse each segment. A time-average ``heat-map" of the numerically simulated 2D wavepacket evolution during the entire gyroscope sequence is given in Fig.~\ref{fig:Gyroscope}D.

In the gyroscope, the potential that we sense is due to a rotating, non-inertial reference frame, and thus is two dimensional, depending on both $x$ and $z$ 
\begin{equation}
    V_I(x, \dot{x}, z, \dot{z};t) = m \Omega \bigl( x \dot{z} - z \dot{x} \bigr)
\end{equation}
where $\Omega$ is the angular velocity of the rotation. Extending the path integral formulation to account for the second dimension gives a design equation for the relative phase accumulated between the clockwise and counterclockwise paths of the gyroscope. It is given by
\begin{equation}
\Delta \theta_{\rm Gy} = \frac{8 \Omega p^2 T_1 T_2}{m \hbar}
\label{eq:gyrophase}
\end{equation}
where the times $T_1$ and $T_2$ are shown in Fig.~\ref{fig:Gyroscope}C. As was the case for the accelerometer, this equation has a simple interpretation of being the rotation parameter $\Omega$ multiplied by twice the enclosed spatial area (twice due to atoms traveling in two opposite directions) in units of~$\hbar/m$. Physically this formula arises from the differential Doppler shifts of the matterwave as it traverses the loops, either in the same or opposite direction to the sense of rotation. The fringe frequency plotted as a function of the rotation angular velocity seen in Fig.~\ref{fig:Gyroscope}F agrees well with the path integral formula in Eq.~(\ref{eq:gyrophase}).

It is worth noting that there is nothing special about the square pattern shown here and other shapes or geometries are possible. For example, one can simply convey the atoms up and down in the $z$ direction using a common translation technique in optical lattice experiments~\cite{PhysRevA.72.053605}, and then apply the matterwave gates only in the $x$ direction. This allows for more flexible patterns including circles that maximize the enclosed area for a given total time. 

\vspace*{-1pc}
\section{Discussion \& Conclusion}
\vspace*{-.75pc}

In this paper, we have presented the design and experimental realization of the matterwave universal gate set capable of constructing a myriad of inertial sensors. The faithful execution of every one of the gates was theoretically demonstrated via numerical simulation and then experimentally confirmed. Experimental validation was carried out by taking successive absorption images and tracking the evolution of the atomic density in the lattice, and by observing their momentum state transformations from time of flight. 

Using the universal gate set, we have demonstrated how to stitch together gates to produce accelerometers, gradiometers, and a gyroscope. Alongside each sensing circuit, we plotted the associated wavepacket evolution from numerical simulation and calculated the expected sensitivity scaling in response to applied signals. Novel split \& hold, conduction-band beamsplitter, and echo gates, offered design solutions that manipulate atoms in a way that is not directly accessible to conventional atom interferometer architectures. 

As presented elsewhere, we have already experimentally demonstrated several of the sensors that we discussed here, including accelerometers in one dimension~\cite{ledesma2023machinedesignedopticallatticeatom} and vector accelerometers in two dimensions~\cite{ledesma2024vector}. Going forward, focus will be on realizing other types of sensor protocols as proposed in this paper. A principal goal will be to measure the sensitivity limits that can be achieved for precision sensing via this method of Bloch-band interferometry.   

Even though this paper focused primarily on one-dimension, we emphasize the straightforward extension of our gates to any or all axes of a 3D optical lattice. This natural extension to multiple dimensions would permit the device to be configured as a multi-axis gyroscope, or even to measure the five independent components of the gradient tensor in a single instrument. Additionally, the flexible programming of Bloch-band gates could enable complex control over lattice-confined atoms for further studies of classical and quantum phase transitions, many-body physics, or the production of novel interferometer geometries. 

We also emphasize that one can learn gates in which the lattice intensity is the control function, or learn gates that incorporate an internal (spin) degree of freedom, opening up many possibilities to extend what we have presented. Finally, the use of entanglement generated by atom interactions would open up a host of new possibilities in leveraging quantum advantage for measurement gains.

\section*{Acknowledgments}
We would like to thank Liang-Ying Chih, Shah Saad Alam and Jarrod Reilly for discussions, and Jieqiu Shao for contributions on quantum optimal control protocols. This research was supported in part by NASA under grant number 80NSSC23K1343, NSF PFC 2317149; NSF OMA 2016244; and NSF PHY 2207963.


\begin{thebibliography}{48}%
\makeatletter
\providecommand \@ifxundefined [1]{%
 \@ifx{#1\undefined}
}%
\providecommand \@ifnum [1]{%
 \ifnum #1\expandafter \@firstoftwo
 \else \expandafter \@secondoftwo
 \fi
}%
\providecommand \@ifx [1]{%
 \ifx #1\expandafter \@firstoftwo
 \else \expandafter \@secondoftwo
 \fi
}%
\providecommand \natexlab [1]{#1}%
\providecommand \enquote  [1]{``#1''}%
\providecommand \bibnamefont  [1]{#1}%
\providecommand \bibfnamefont [1]{#1}%
\providecommand \citenamefont [1]{#1}%
\providecommand \href@noop [0]{\@secondoftwo}%
\providecommand \href [0]{\begingroup \@sanitize@url \@href}%
\providecommand \@href[1]{\@@startlink{#1}\@@href}%
\providecommand \@@href[1]{\endgroup#1\@@endlink}%
\providecommand \@sanitize@url [0]{\catcode `\\12\catcode `\$12\catcode `\&12\catcode `\#12\catcode `\^12\catcode `\_12\catcode `\%12\relax}%
\providecommand \@@startlink[1]{}%
\providecommand \@@endlink[0]{}%
\providecommand \url  [0]{\begingroup\@sanitize@url \@url }%
\providecommand \@url [1]{\endgroup\@href {#1}{\urlprefix }}%
\providecommand \urlprefix  [0]{URL }%
\providecommand \Eprint [0]{\href }%
\providecommand \doibase [0]{https://doi.org/}%
\providecommand \selectlanguage [0]{\@gobble}%
\providecommand \bibinfo  [0]{\@secondoftwo}%
\providecommand \bibfield  [0]{\@secondoftwo}%
\providecommand \translation [1]{[#1]}%
\providecommand \BibitemOpen [0]{}%
\providecommand \bibitemStop [0]{}%
\providecommand \bibitemNoStop [0]{.\EOS\space}%
\providecommand \EOS [0]{\spacefactor3000\relax}%
\providecommand \BibitemShut  [1]{\csname bibitem#1\endcsname}%
\let\auto@bib@innerbib\@empty
\bibitem [{\citenamefont {Takamoto}\ \emph {et~al.}(2005)\citenamefont {Takamoto}, \citenamefont {Hong}, \citenamefont {Higashi},\ and\ \citenamefont {Katori}}]{Takamoto2005}%
  \BibitemOpen
  \bibfield  {author} {\bibinfo {author} {\bibfnamefont {M.}~\bibnamefont {Takamoto}}, \bibinfo {author} {\bibfnamefont {F.-L.}\ \bibnamefont {Hong}}, \bibinfo {author} {\bibfnamefont {R.}~\bibnamefont {Higashi}},\ and\ \bibinfo {author} {\bibfnamefont {H.}~\bibnamefont {Katori}},\ }\bibfield  {title} {\bibinfo {title} {An optical lattice clock},\ }\href {https://doi.org/10.1038/nature03541} {\bibfield  {journal} {\bibinfo  {journal} {Nature}\ }\textbf {\bibinfo {volume} {435}},\ \bibinfo {pages} {321} (\bibinfo {year} {2005})}\BibitemShut {NoStop}%
\bibitem [{\citenamefont {Kim}\ \emph {et~al.}(2023)\citenamefont {Kim}, \citenamefont {Aeppli}, \citenamefont {Bothwell},\ and\ \citenamefont {Ye}}]{PhysRevLett.130.113203}%
  \BibitemOpen
  \bibfield  {author} {\bibinfo {author} {\bibfnamefont {K.}~\bibnamefont {Kim}}, \bibinfo {author} {\bibfnamefont {A.}~\bibnamefont {Aeppli}}, \bibinfo {author} {\bibfnamefont {T.}~\bibnamefont {Bothwell}},\ and\ \bibinfo {author} {\bibfnamefont {J.}~\bibnamefont {Ye}},\ }\bibfield  {title} {\bibinfo {title} {Evaluation of lattice light shift at low ${10}^{\ensuremath{-}19}$ uncertainty for a shallow lattice sr optical clock},\ }\href {https://doi.org/10.1103/PhysRevLett.130.113203} {\bibfield  {journal} {\bibinfo  {journal} {Phys. Rev. Lett.}\ }\textbf {\bibinfo {volume} {130}},\ \bibinfo {pages} {113203} (\bibinfo {year} {2023})}\BibitemShut {NoStop}%
\bibitem [{\citenamefont {Keith}\ \emph {et~al.}(1991)\citenamefont {Keith}, \citenamefont {Ekstrom}, \citenamefont {Turchette},\ and\ \citenamefont {Pritchard}}]{PhysRevLett.66.2693}%
  \BibitemOpen
  \bibfield  {author} {\bibinfo {author} {\bibfnamefont {D.~W.}\ \bibnamefont {Keith}}, \bibinfo {author} {\bibfnamefont {C.~R.}\ \bibnamefont {Ekstrom}}, \bibinfo {author} {\bibfnamefont {Q.~A.}\ \bibnamefont {Turchette}},\ and\ \bibinfo {author} {\bibfnamefont {D.~E.}\ \bibnamefont {Pritchard}},\ }\bibfield  {title} {\bibinfo {title} {An interferometer for atoms},\ }\href {https://doi.org/10.1103/PhysRevLett.66.2693} {\bibfield  {journal} {\bibinfo  {journal} {Phys. Rev. Lett.}\ }\textbf {\bibinfo {volume} {66}},\ \bibinfo {pages} {2693} (\bibinfo {year} {1991})}\BibitemShut {NoStop}%
\bibitem [{\citenamefont {Kasevich}\ and\ \citenamefont {Chu}(1991)}]{PhysRevLett.67.181}%
  \BibitemOpen
  \bibfield  {author} {\bibinfo {author} {\bibfnamefont {M.}~\bibnamefont {Kasevich}}\ and\ \bibinfo {author} {\bibfnamefont {S.}~\bibnamefont {Chu}},\ }\bibfield  {title} {\bibinfo {title} {Atomic interferometry using stimulated raman transitions},\ }\href {https://doi.org/10.1103/PhysRevLett.67.181} {\bibfield  {journal} {\bibinfo  {journal} {Phys. Rev. Lett.}\ }\textbf {\bibinfo {volume} {67}},\ \bibinfo {pages} {181} (\bibinfo {year} {1991})}\BibitemShut {NoStop}%
\bibitem [{\citenamefont {Kasevich}\ and\ \citenamefont {Chu}(1992)}]{LightPulse_Kasevich}%
  \BibitemOpen
  \bibfield  {author} {\bibinfo {author} {\bibfnamefont {M.}~\bibnamefont {Kasevich}}\ and\ \bibinfo {author} {\bibfnamefont {S.}~\bibnamefont {Chu}},\ }\bibfield  {title} {\bibinfo {title} {Measurement of the gravitational acceleration of an atom with a light-pulse atom interferometer},\ }\href@noop {} {\bibfield  {journal} {\bibinfo  {journal} {Applied Physics B}\ }\textbf {\bibinfo {volume} {54}},\ \bibinfo {pages} {321} (\bibinfo {year} {1992})}\BibitemShut {NoStop}%
\bibitem [{\citenamefont {Gustavson}\ \emph {et~al.}(1997)\citenamefont {Gustavson}, \citenamefont {Bouyer},\ and\ \citenamefont {Kasevich}}]{PhysRevLett.78.2046}%
  \BibitemOpen
  \bibfield  {author} {\bibinfo {author} {\bibfnamefont {T.~L.}\ \bibnamefont {Gustavson}}, \bibinfo {author} {\bibfnamefont {P.}~\bibnamefont {Bouyer}},\ and\ \bibinfo {author} {\bibfnamefont {M.~A.}\ \bibnamefont {Kasevich}},\ }\bibfield  {title} {\bibinfo {title} {Precision rotation measurements with an atom interferometer gyroscope},\ }\href {https://doi.org/10.1103/PhysRevLett.78.2046} {\bibfield  {journal} {\bibinfo  {journal} {Phys. Rev. Lett.}\ }\textbf {\bibinfo {volume} {78}},\ \bibinfo {pages} {2046} (\bibinfo {year} {1997})}\BibitemShut {NoStop}%
\bibitem [{\citenamefont {Fang}\ and\ \citenamefont {Qin}(2012)}]{s120506331}%
  \BibitemOpen
  \bibfield  {author} {\bibinfo {author} {\bibfnamefont {J.}~\bibnamefont {Fang}}\ and\ \bibinfo {author} {\bibfnamefont {J.}~\bibnamefont {Qin}},\ }\bibfield  {title} {\bibinfo {title} {Advances in atomic gyroscopes: A view from inertial navigation applications},\ }\href {https://doi.org/10.3390/s120506331} {\bibfield  {journal} {\bibinfo  {journal} {Sensors}\ }\textbf {\bibinfo {volume} {12}},\ \bibinfo {pages} {6331} (\bibinfo {year} {2012})}\BibitemShut {NoStop}%
\bibitem [{\citenamefont {Snadden}\ \emph {et~al.}(1998)\citenamefont {Snadden}, \citenamefont {McGuirk}, \citenamefont {Bouyer}, \citenamefont {Haritos},\ and\ \citenamefont {Kasevich}}]{PhysRevLett.81.971}%
  \BibitemOpen
  \bibfield  {author} {\bibinfo {author} {\bibfnamefont {M.~J.}\ \bibnamefont {Snadden}}, \bibinfo {author} {\bibfnamefont {J.~M.}\ \bibnamefont {McGuirk}}, \bibinfo {author} {\bibfnamefont {P.}~\bibnamefont {Bouyer}}, \bibinfo {author} {\bibfnamefont {K.~G.}\ \bibnamefont {Haritos}},\ and\ \bibinfo {author} {\bibfnamefont {M.~A.}\ \bibnamefont {Kasevich}},\ }\bibfield  {title} {\bibinfo {title} {Measurement of the earth's gravity gradient with an atom interferometer-based gravity gradiometer},\ }\href {https://doi.org/10.1103/PhysRevLett.81.971} {\bibfield  {journal} {\bibinfo  {journal} {Phys. Rev. Lett.}\ }\textbf {\bibinfo {volume} {81}},\ \bibinfo {pages} {971} (\bibinfo {year} {1998})}\BibitemShut {NoStop}%
\bibitem [{\citenamefont {Bertoldi}\ \emph {et~al.}(2006)\citenamefont {Bertoldi}, \citenamefont {Lamporesi}, \citenamefont {Cacciapuoti}, \citenamefont {de~Angelis}, \citenamefont {Fattori}, \citenamefont {Petelski}, \citenamefont {Peters}, \citenamefont {Prevedelli}, \citenamefont {Stuhler},\ and\ \citenamefont {Tino}}]{Bertoldi2006}%
  \BibitemOpen
  \bibfield  {author} {\bibinfo {author} {\bibfnamefont {A.}~\bibnamefont {Bertoldi}}, \bibinfo {author} {\bibfnamefont {G.}~\bibnamefont {Lamporesi}}, \bibinfo {author} {\bibfnamefont {L.}~\bibnamefont {Cacciapuoti}}, \bibinfo {author} {\bibfnamefont {M.}~\bibnamefont {de~Angelis}}, \bibinfo {author} {\bibfnamefont {M.}~\bibnamefont {Fattori}}, \bibinfo {author} {\bibfnamefont {T.}~\bibnamefont {Petelski}}, \bibinfo {author} {\bibfnamefont {A.}~\bibnamefont {Peters}}, \bibinfo {author} {\bibfnamefont {M.}~\bibnamefont {Prevedelli}}, \bibinfo {author} {\bibfnamefont {J.}~\bibnamefont {Stuhler}},\ and\ \bibinfo {author} {\bibfnamefont {G.~M.}\ \bibnamefont {Tino}},\ }\bibfield  {title} {\bibinfo {title} {Atom interferometry gravity-gradiometer for the determination of the newtonian gravitational constant g},\ }\href {https://doi.org/10.1140/epjd/e2006-00212-2} {\bibfield  {journal} {\bibinfo  {journal} {The European Physical Journal D - Atomic, Molecular, Optical and Plasma Physics}\ }\textbf {\bibinfo
  {volume} {40}},\ \bibinfo {pages} {271} (\bibinfo {year} {2006})}\BibitemShut {NoStop}%
\bibitem [{\citenamefont {Weiss}\ \emph {et~al.}(1993)\citenamefont {Weiss}, \citenamefont {Young},\ and\ \citenamefont {Chu}}]{PhysRevLett.70.2706}%
  \BibitemOpen
  \bibfield  {author} {\bibinfo {author} {\bibfnamefont {D.~S.}\ \bibnamefont {Weiss}}, \bibinfo {author} {\bibfnamefont {B.~C.}\ \bibnamefont {Young}},\ and\ \bibinfo {author} {\bibfnamefont {S.}~\bibnamefont {Chu}},\ }\bibfield  {title} {\bibinfo {title} {Precision measurement of the photon recoil of an atom using atomic interferometry},\ }\href {https://doi.org/10.1103/PhysRevLett.70.2706} {\bibfield  {journal} {\bibinfo  {journal} {Phys. Rev. Lett.}\ }\textbf {\bibinfo {volume} {70}},\ \bibinfo {pages} {2706} (\bibinfo {year} {1993})}\BibitemShut {NoStop}%
\bibitem [{\citenamefont {Lamporesi}\ \emph {et~al.}(2008)\citenamefont {Lamporesi}, \citenamefont {Bertoldi}, \citenamefont {Cacciapuoti}, \citenamefont {Prevedelli},\ and\ \citenamefont {Tino}}]{PhysRevLett.100.050801}%
  \BibitemOpen
  \bibfield  {author} {\bibinfo {author} {\bibfnamefont {G.}~\bibnamefont {Lamporesi}}, \bibinfo {author} {\bibfnamefont {A.}~\bibnamefont {Bertoldi}}, \bibinfo {author} {\bibfnamefont {L.}~\bibnamefont {Cacciapuoti}}, \bibinfo {author} {\bibfnamefont {M.}~\bibnamefont {Prevedelli}},\ and\ \bibinfo {author} {\bibfnamefont {G.~M.}\ \bibnamefont {Tino}},\ }\bibfield  {title} {\bibinfo {title} {Determination of the newtonian gravitational constant using atom interferometry},\ }\href {https://doi.org/10.1103/PhysRevLett.100.050801} {\bibfield  {journal} {\bibinfo  {journal} {Phys. Rev. Lett.}\ }\textbf {\bibinfo {volume} {100}},\ \bibinfo {pages} {050801} (\bibinfo {year} {2008})}\BibitemShut {NoStop}%
\bibitem [{\citenamefont {Hamilton}\ \emph {et~al.}(2015{\natexlab{a}})\citenamefont {Hamilton}, \citenamefont {Jaffe}, \citenamefont {Haslinger}, \citenamefont {Simmons}, \citenamefont {Muller},\ and\ \citenamefont {Khoury}}]{doi:10.1126/science.aaa8883}%
  \BibitemOpen
  \bibfield  {author} {\bibinfo {author} {\bibfnamefont {P.}~\bibnamefont {Hamilton}}, \bibinfo {author} {\bibfnamefont {M.}~\bibnamefont {Jaffe}}, \bibinfo {author} {\bibfnamefont {P.}~\bibnamefont {Haslinger}}, \bibinfo {author} {\bibfnamefont {Q.}~\bibnamefont {Simmons}}, \bibinfo {author} {\bibfnamefont {H.}~\bibnamefont {Muller}},\ and\ \bibinfo {author} {\bibfnamefont {J.}~\bibnamefont {Khoury}},\ }\bibfield  {title} {\bibinfo {title} {Atom-interferometry constraints on dark energy},\ }\href {https://doi.org/10.1126/science.aaa8883} {\bibfield  {journal} {\bibinfo  {journal} {Science}\ }\textbf {\bibinfo {volume} {349}},\ \bibinfo {pages} {849} (\bibinfo {year} {2015}{\natexlab{a}})}\BibitemShut {NoStop}%
\bibitem [{\citenamefont {Asenbaum}\ \emph {et~al.}(2020)\citenamefont {Asenbaum}, \citenamefont {Overstreet}, \citenamefont {Kim}, \citenamefont {Curti},\ and\ \citenamefont {Kasevich}}]{EquivPrinc_Kasevich}%
  \BibitemOpen
  \bibfield  {author} {\bibinfo {author} {\bibfnamefont {P.}~\bibnamefont {Asenbaum}}, \bibinfo {author} {\bibfnamefont {C.}~\bibnamefont {Overstreet}}, \bibinfo {author} {\bibfnamefont {M.}~\bibnamefont {Kim}}, \bibinfo {author} {\bibfnamefont {J.}~\bibnamefont {Curti}},\ and\ \bibinfo {author} {\bibfnamefont {M.~A.}\ \bibnamefont {Kasevich}},\ }\bibfield  {title} {\bibinfo {title} {Atom-interferometric test of the equivalence principle at the ${10}^{\ensuremath{-}12}$ level},\ }\href {https://doi.org/10.1103/PhysRevLett.125.191101} {\bibfield  {journal} {\bibinfo  {journal} {Phys. Rev. Lett.}\ }\textbf {\bibinfo {volume} {125}},\ \bibinfo {pages} {191101} (\bibinfo {year} {2020})}\BibitemShut {NoStop}%
\bibitem [{\citenamefont {Herrmann}\ \emph {et~al.}(2012)\citenamefont {Herrmann}, \citenamefont {Dittus}, \citenamefont {Lämmerzahl}, \citenamefont {(for~the QUANTUS},\ and\ \citenamefont {teams)}}]{Herrmann_2012}%
  \BibitemOpen
  \bibfield  {author} {\bibinfo {author} {\bibfnamefont {S.}~\bibnamefont {Herrmann}}, \bibinfo {author} {\bibfnamefont {H.}~\bibnamefont {Dittus}}, \bibinfo {author} {\bibfnamefont {C.}~\bibnamefont {Lämmerzahl}}, \bibinfo {author} {\bibnamefont {(for~the QUANTUS}},\ and\ \bibinfo {author} {\bibfnamefont {P.}~\bibnamefont {teams)}},\ }\bibfield  {title} {\bibinfo {title} {Testing the equivalence principle with atomic interferometry},\ }\href {https://doi.org/10.1088/0264-9381/29/18/184003} {\bibfield  {journal} {\bibinfo  {journal} {Classical and Quantum Gravity}\ }\textbf {\bibinfo {volume} {29}},\ \bibinfo {pages} {184003} (\bibinfo {year} {2012})}\BibitemShut {NoStop}%
\bibitem [{\citenamefont {Barrett}\ \emph {et~al.}(2019)\citenamefont {Barrett}, \citenamefont {Cheiney}, \citenamefont {Battelier}, \citenamefont {Napolitano},\ and\ \citenamefont {Bouyer}}]{PhysRevLett.122.043604}%
  \BibitemOpen
  \bibfield  {author} {\bibinfo {author} {\bibfnamefont {B.}~\bibnamefont {Barrett}}, \bibinfo {author} {\bibfnamefont {P.}~\bibnamefont {Cheiney}}, \bibinfo {author} {\bibfnamefont {B.}~\bibnamefont {Battelier}}, \bibinfo {author} {\bibfnamefont {F.}~\bibnamefont {Napolitano}},\ and\ \bibinfo {author} {\bibfnamefont {P.}~\bibnamefont {Bouyer}},\ }\bibfield  {title} {\bibinfo {title} {Multidimensional atom optics and interferometry},\ }\href {https://doi.org/10.1103/PhysRevLett.122.043604} {\bibfield  {journal} {\bibinfo  {journal} {Phys. Rev. Lett.}\ }\textbf {\bibinfo {volume} {122}},\ \bibinfo {pages} {043604} (\bibinfo {year} {2019})}\BibitemShut {NoStop}%
\bibitem [{\citenamefont {Savoie}\ \emph {et~al.}(2018)\citenamefont {Savoie}, \citenamefont {Altorio}, \citenamefont {Fang}, \citenamefont {Sidorenkov}, \citenamefont {Geiger},\ and\ \citenamefont {Landragin}}]{doi:10.1126/sciadv.aau7948}%
  \BibitemOpen
  \bibfield  {author} {\bibinfo {author} {\bibfnamefont {D.}~\bibnamefont {Savoie}}, \bibinfo {author} {\bibfnamefont {M.}~\bibnamefont {Altorio}}, \bibinfo {author} {\bibfnamefont {B.}~\bibnamefont {Fang}}, \bibinfo {author} {\bibfnamefont {L.~A.}\ \bibnamefont {Sidorenkov}}, \bibinfo {author} {\bibfnamefont {R.}~\bibnamefont {Geiger}},\ and\ \bibinfo {author} {\bibfnamefont {A.}~\bibnamefont {Landragin}},\ }\bibfield  {title} {\bibinfo {title} {Interleaved atom interferometry for high-sensitivity inertial measurements},\ }\href {https://doi.org/10.1126/sciadv.aau7948} {\bibfield  {journal} {\bibinfo  {journal} {Science Advances}\ }\textbf {\bibinfo {volume} {4}},\ \bibinfo {pages} {eaau7948} (\bibinfo {year} {2018})}\BibitemShut {NoStop}%
\bibitem [{\citenamefont {Menoret}\ \emph {et~al.}(2018)\citenamefont {Menoret}, \citenamefont {Vermeulen}, \citenamefont {Le~Moigne}, \citenamefont {Bonvalot}, \citenamefont {Bouyer}, \citenamefont {Landragin},\ and\ \citenamefont {Desruelle}}]{Ménoret2018}%
  \BibitemOpen
  \bibfield  {author} {\bibinfo {author} {\bibfnamefont {V.}~\bibnamefont {Menoret}}, \bibinfo {author} {\bibfnamefont {P.}~\bibnamefont {Vermeulen}}, \bibinfo {author} {\bibfnamefont {N.}~\bibnamefont {Le~Moigne}}, \bibinfo {author} {\bibfnamefont {S.}~\bibnamefont {Bonvalot}}, \bibinfo {author} {\bibfnamefont {P.}~\bibnamefont {Bouyer}}, \bibinfo {author} {\bibfnamefont {A.}~\bibnamefont {Landragin}},\ and\ \bibinfo {author} {\bibfnamefont {B.}~\bibnamefont {Desruelle}},\ }\bibfield  {title} {\bibinfo {title} {Gravity measurements below 10-9 g with a transportable absolute quantum gravimeter},\ }\href {https://doi.org/10.1038/s41598-018-30608-1} {\bibfield  {journal} {\bibinfo  {journal} {Scientific Reports}\ }\textbf {\bibinfo {volume} {8}},\ \bibinfo {pages} {12300} (\bibinfo {year} {2018})}\BibitemShut {NoStop}%
\bibitem [{\citenamefont {Lee}\ \emph {et~al.}(2022)\citenamefont {Lee}, \citenamefont {Ding}, \citenamefont {Christensen}, \citenamefont {Rosenthal}, \citenamefont {Ison}, \citenamefont {Gillund}, \citenamefont {Bossert}, \citenamefont {Fuerschbach}, \citenamefont {Kindel}, \citenamefont {Finnegan}, \citenamefont {Wendt}, \citenamefont {Gehl}, \citenamefont {Kodigala}, \citenamefont {McGuinness}, \citenamefont {Walker}, \citenamefont {Kemme}, \citenamefont {Lentine}, \citenamefont {Biedermann},\ and\ \citenamefont {Schwindt}}]{Lee2022}%
  \BibitemOpen
  \bibfield  {author} {\bibinfo {author} {\bibfnamefont {J.}~\bibnamefont {Lee}}, \bibinfo {author} {\bibfnamefont {R.}~\bibnamefont {Ding}}, \bibinfo {author} {\bibfnamefont {J.}~\bibnamefont {Christensen}}, \bibinfo {author} {\bibfnamefont {R.~R.}\ \bibnamefont {Rosenthal}}, \bibinfo {author} {\bibfnamefont {A.}~\bibnamefont {Ison}}, \bibinfo {author} {\bibfnamefont {D.~P.}\ \bibnamefont {Gillund}}, \bibinfo {author} {\bibfnamefont {D.}~\bibnamefont {Bossert}}, \bibinfo {author} {\bibfnamefont {K.~H.}\ \bibnamefont {Fuerschbach}}, \bibinfo {author} {\bibfnamefont {W.}~\bibnamefont {Kindel}}, \bibinfo {author} {\bibfnamefont {P.~S.}\ \bibnamefont {Finnegan}}, \bibinfo {author} {\bibfnamefont {J.~R.}\ \bibnamefont {Wendt}}, \bibinfo {author} {\bibfnamefont {M.}~\bibnamefont {Gehl}}, \bibinfo {author} {\bibfnamefont {A.}~\bibnamefont {Kodigala}}, \bibinfo {author} {\bibfnamefont {H.}~\bibnamefont {McGuinness}}, \bibinfo {author} {\bibfnamefont {C.~A.}\ \bibnamefont {Walker}}, \bibinfo {author} {\bibfnamefont
  {S.~A.}\ \bibnamefont {Kemme}}, \bibinfo {author} {\bibfnamefont {A.}~\bibnamefont {Lentine}}, \bibinfo {author} {\bibfnamefont {G.}~\bibnamefont {Biedermann}},\ and\ \bibinfo {author} {\bibfnamefont {P.~D.~D.}\ \bibnamefont {Schwindt}},\ }\bibfield  {title} {\bibinfo {title} {A compact cold-atom interferometer with a high data-rate grating magneto-optical trap and a photonic-integrated-circuit-compatible laser system},\ }\href {https://doi.org/10.1038/s41467-022-31410-4} {\bibfield  {journal} {\bibinfo  {journal} {Nature Communications}\ }\textbf {\bibinfo {volume} {13}},\ \bibinfo {pages} {5131} (\bibinfo {year} {2022})}\BibitemShut {NoStop}%
\bibitem [{\citenamefont {LeDesma}\ \emph {et~al.}(2024{\natexlab{a}})\citenamefont {LeDesma}, \citenamefont {Mehling},\ and\ \citenamefont {Holland}}]{ledesma2024vector}%
  \BibitemOpen
  \bibfield  {author} {\bibinfo {author} {\bibfnamefont {C.}~\bibnamefont {LeDesma}}, \bibinfo {author} {\bibfnamefont {K.}~\bibnamefont {Mehling}},\ and\ \bibinfo {author} {\bibfnamefont {M.}~\bibnamefont {Holland}},\ }\bibfield  {title} {\bibinfo {title} {Vector atom accelerometry in an optical lattice},\ }\href@noop {} {\bibfield  {journal} {\bibinfo  {journal} {arXiv preprint arXiv:2407.04874}\ } (\bibinfo {year} {2024}{\natexlab{a}})}\BibitemShut {NoStop}%
\bibitem [{\citenamefont {Feynman}(1986)}]{Feynman1986}%
  \BibitemOpen
  \bibfield  {author} {\bibinfo {author} {\bibfnamefont {R.~P.}\ \bibnamefont {Feynman}},\ }\bibfield  {title} {\bibinfo {title} {Quantum mechanical computers},\ }\href {https://doi.org/10.1007/BF01886518} {\bibfield  {journal} {\bibinfo  {journal} {Foundations of Physics}\ }\textbf {\bibinfo {volume} {16}},\ \bibinfo {pages} {507} (\bibinfo {year} {1986})}\BibitemShut {NoStop}%
\bibitem [{\citenamefont {DiVincenzo}(1995)}]{PhysRevA.51.1015}%
  \BibitemOpen
  \bibfield  {author} {\bibinfo {author} {\bibfnamefont {D.~P.}\ \bibnamefont {DiVincenzo}},\ }\bibfield  {title} {\bibinfo {title} {Two-bit gates are universal for quantum computation},\ }\href {https://doi.org/10.1103/PhysRevA.51.1015} {\bibfield  {journal} {\bibinfo  {journal} {Phys. Rev. A}\ }\textbf {\bibinfo {volume} {51}},\ \bibinfo {pages} {1015} (\bibinfo {year} {1995})}\BibitemShut {NoStop}%
\bibitem [{\citenamefont {Potting}\ \emph {et~al.}(2001)\citenamefont {Potting}, \citenamefont {Cramer},\ and\ \citenamefont {Meystre}}]{PhysRevA.64.063613}%
  \BibitemOpen
  \bibfield  {author} {\bibinfo {author} {\bibfnamefont {S.}~\bibnamefont {Potting}}, \bibinfo {author} {\bibfnamefont {M.}~\bibnamefont {Cramer}},\ and\ \bibinfo {author} {\bibfnamefont {P.}~\bibnamefont {Meystre}},\ }\bibfield  {title} {\bibinfo {title} {Momentum-state engineering and control in bose-einstein condensates},\ }\href {https://doi.org/10.1103/PhysRevA.64.063613} {\bibfield  {journal} {\bibinfo  {journal} {Phys. Rev. A}\ }\textbf {\bibinfo {volume} {64}},\ \bibinfo {pages} {063613} (\bibinfo {year} {2001})}\BibitemShut {NoStop}%
\bibitem [{\citenamefont {Dupont}\ \emph {et~al.}(2021)\citenamefont {Dupont}, \citenamefont {Chatelain}, \citenamefont {Gabardos}, \citenamefont {Arnal}, \citenamefont {Billy}, \citenamefont {Peaudecerf}, \citenamefont {Sugny},\ and\ \citenamefont {Gu\'ery-Odelin}}]{PRXQuantum.2.040303}%
  \BibitemOpen
  \bibfield  {author} {\bibinfo {author} {\bibfnamefont {N.}~\bibnamefont {Dupont}}, \bibinfo {author} {\bibfnamefont {G.}~\bibnamefont {Chatelain}}, \bibinfo {author} {\bibfnamefont {L.}~\bibnamefont {Gabardos}}, \bibinfo {author} {\bibfnamefont {M.}~\bibnamefont {Arnal}}, \bibinfo {author} {\bibfnamefont {J.}~\bibnamefont {Billy}}, \bibinfo {author} {\bibfnamefont {B.}~\bibnamefont {Peaudecerf}}, \bibinfo {author} {\bibfnamefont {D.}~\bibnamefont {Sugny}},\ and\ \bibinfo {author} {\bibfnamefont {D.}~\bibnamefont {Gu\'ery-Odelin}},\ }\bibfield  {title} {\bibinfo {title} {Quantum state control of a bose-einstein condensate in an optical lattice},\ }\href {https://doi.org/10.1103/PRXQuantum.2.040303} {\bibfield  {journal} {\bibinfo  {journal} {PRX Quantum}\ }\textbf {\bibinfo {volume} {2}},\ \bibinfo {pages} {040303} (\bibinfo {year} {2021})}\BibitemShut {NoStop}%
\bibitem [{\citenamefont {Nicotra}\ \emph {et~al.}(2023)\citenamefont {Nicotra}, \citenamefont {Shao}, \citenamefont {Combes}, \citenamefont {Theurkauf}, \citenamefont {Axelrad}, \citenamefont {Chih}, \citenamefont {Holland}, \citenamefont {Zozulya}, \citenamefont {LeDesma}, \citenamefont {Mehling},\ and\ \citenamefont {Anderson}}]{10015539}%
  \BibitemOpen
  \bibfield  {author} {\bibinfo {author} {\bibfnamefont {M.~M.}\ \bibnamefont {Nicotra}}, \bibinfo {author} {\bibfnamefont {J.}~\bibnamefont {Shao}}, \bibinfo {author} {\bibfnamefont {J.}~\bibnamefont {Combes}}, \bibinfo {author} {\bibfnamefont {A.~C.}\ \bibnamefont {Theurkauf}}, \bibinfo {author} {\bibfnamefont {P.}~\bibnamefont {Axelrad}}, \bibinfo {author} {\bibfnamefont {L.-Y.}\ \bibnamefont {Chih}}, \bibinfo {author} {\bibfnamefont {M.}~\bibnamefont {Holland}}, \bibinfo {author} {\bibfnamefont {A.~A.}\ \bibnamefont {Zozulya}}, \bibinfo {author} {\bibfnamefont {C.~K.}\ \bibnamefont {LeDesma}}, \bibinfo {author} {\bibfnamefont {K.}~\bibnamefont {Mehling}},\ and\ \bibinfo {author} {\bibfnamefont {D.~Z.}\ \bibnamefont {Anderson}},\ }\bibfield  {title} {\bibinfo {title} {Modeling and control of ultracold atoms trapped in an optical lattice: An example-driven tutorial on quantum control},\ }\href {https://doi.org/10.1109/MCS.2022.3216652} {\bibfield  {journal} {\bibinfo  {journal} {IEEE Control Systems
  Magazine}\ }\textbf {\bibinfo {volume} {43}},\ \bibinfo {pages} {28} (\bibinfo {year} {2023})}\BibitemShut {NoStop}%
\bibitem [{\citenamefont {Madison}\ \emph {et~al.}(1998)\citenamefont {Madison}, \citenamefont {Fischer}, \citenamefont {Diener}, \citenamefont {Niu},\ and\ \citenamefont {Raizen}}]{PhysRevLett.81.5093}%
  \BibitemOpen
  \bibfield  {author} {\bibinfo {author} {\bibfnamefont {K.~W.}\ \bibnamefont {Madison}}, \bibinfo {author} {\bibfnamefont {M.~C.}\ \bibnamefont {Fischer}}, \bibinfo {author} {\bibfnamefont {R.~B.}\ \bibnamefont {Diener}}, \bibinfo {author} {\bibfnamefont {Q.}~\bibnamefont {Niu}},\ and\ \bibinfo {author} {\bibfnamefont {M.~G.}\ \bibnamefont {Raizen}},\ }\bibfield  {title} {\bibinfo {title} {Dynamical bloch band suppression in an optical lattice},\ }\href {https://doi.org/10.1103/PhysRevLett.81.5093} {\bibfield  {journal} {\bibinfo  {journal} {Phys. Rev. Lett.}\ }\textbf {\bibinfo {volume} {81}},\ \bibinfo {pages} {5093} (\bibinfo {year} {1998})}\BibitemShut {NoStop}%
\bibitem [{\citenamefont {Schiavoni}\ \emph {et~al.}(2003)\citenamefont {Schiavoni}, \citenamefont {Sanchez-Palencia}, \citenamefont {Renzoni},\ and\ \citenamefont {Grynberg}}]{PhysRevLett.90.094101}%
  \BibitemOpen
  \bibfield  {author} {\bibinfo {author} {\bibfnamefont {M.}~\bibnamefont {Schiavoni}}, \bibinfo {author} {\bibfnamefont {L.}~\bibnamefont {Sanchez-Palencia}}, \bibinfo {author} {\bibfnamefont {F.}~\bibnamefont {Renzoni}},\ and\ \bibinfo {author} {\bibfnamefont {G.}~\bibnamefont {Grynberg}},\ }\bibfield  {title} {\bibinfo {title} {Phase control of directed diffusion in a symmetric optical lattice},\ }\href {https://doi.org/10.1103/PhysRevLett.90.094101} {\bibfield  {journal} {\bibinfo  {journal} {Phys. Rev. Lett.}\ }\textbf {\bibinfo {volume} {90}},\ \bibinfo {pages} {094101} (\bibinfo {year} {2003})}\BibitemShut {NoStop}%
\bibitem [{\citenamefont {Denschlag}\ \emph {et~al.}(2002)\citenamefont {Denschlag}, \citenamefont {Simsarian}, \citenamefont {Haffner}, \citenamefont {McKenzie}, \citenamefont {Browaeys}, \citenamefont {Cho}, \citenamefont {Helmerson}, \citenamefont {Rolston},\ and\ \citenamefont {Phillips}}]{Denschlag2002}%
  \BibitemOpen
  \bibfield  {author} {\bibinfo {author} {\bibfnamefont {J.~H.}\ \bibnamefont {Denschlag}}, \bibinfo {author} {\bibfnamefont {J.~E.}\ \bibnamefont {Simsarian}}, \bibinfo {author} {\bibfnamefont {H.}~\bibnamefont {Haffner}}, \bibinfo {author} {\bibfnamefont {C.}~\bibnamefont {McKenzie}}, \bibinfo {author} {\bibfnamefont {A.}~\bibnamefont {Browaeys}}, \bibinfo {author} {\bibfnamefont {D.}~\bibnamefont {Cho}}, \bibinfo {author} {\bibfnamefont {K.}~\bibnamefont {Helmerson}}, \bibinfo {author} {\bibfnamefont {S.~L.}\ \bibnamefont {Rolston}},\ and\ \bibinfo {author} {\bibfnamefont {W.~D.}\ \bibnamefont {Phillips}},\ }\bibfield  {title} {\bibinfo {title} {A bose-einstein condensate in an optical lattice},\ }\href {https://doi.org/10.1088/0953-4075/35/14/307} {\bibfield  {journal} {\bibinfo  {journal} {Journal of Physics B: Atomic, Molecular and Optical Physics}\ }\textbf {\bibinfo {volume} {35}},\ \bibinfo {pages} {3095} (\bibinfo {year} {2002})}\BibitemShut {NoStop}%
\bibitem [{\citenamefont {Chih}\ and\ \citenamefont {Holland}(2021)}]{RLMatter_Chih}%
  \BibitemOpen
  \bibfield  {author} {\bibinfo {author} {\bibfnamefont {L.-Y.}\ \bibnamefont {Chih}}\ and\ \bibinfo {author} {\bibfnamefont {M.}~\bibnamefont {Holland}},\ }\bibfield  {title} {\bibinfo {title} {Reinforcement-learning-based matter-wave interferometer in a shaken optical lattice},\ }\href {https://doi.org/10.1103/PhysRevResearch.3.033279} {\bibfield  {journal} {\bibinfo  {journal} {Phys. Rev. Research}\ }\textbf {\bibinfo {volume} {3}},\ \bibinfo {pages} {033279} (\bibinfo {year} {2021})}\BibitemShut {NoStop}%
\bibitem [{\citenamefont {LeDesma}\ \emph {et~al.}(2023)\citenamefont {LeDesma}, \citenamefont {Mehling}, \citenamefont {Shao}, \citenamefont {Wilson}, \citenamefont {Axelrad}, \citenamefont {Nicotra}, \citenamefont {Holland},\ and\ \citenamefont {Anderson}}]{ledesma2023machinedesignedopticallatticeatom}%
  \BibitemOpen
  \bibfield  {author} {\bibinfo {author} {\bibfnamefont {C.}~\bibnamefont {LeDesma}}, \bibinfo {author} {\bibfnamefont {K.}~\bibnamefont {Mehling}}, \bibinfo {author} {\bibfnamefont {J.}~\bibnamefont {Shao}}, \bibinfo {author} {\bibfnamefont {J.~D.}\ \bibnamefont {Wilson}}, \bibinfo {author} {\bibfnamefont {P.}~\bibnamefont {Axelrad}}, \bibinfo {author} {\bibfnamefont {M.~M.}\ \bibnamefont {Nicotra}}, \bibinfo {author} {\bibfnamefont {M.}~\bibnamefont {Holland}},\ and\ \bibinfo {author} {\bibfnamefont {D.~Z.}\ \bibnamefont {Anderson}},\ }\href {https://arxiv.org/abs/2305.17603} {\bibinfo {title} {A machine-designed optical lattice atom interferometer}} (\bibinfo {year} {2023}),\ \Eprint {https://arxiv.org/abs/2305.17603} {arXiv:2305.17603 [quant-ph]} \BibitemShut {NoStop}%
\bibitem [{\citenamefont {LeDesma}\ \emph {et~al.}(2024{\natexlab{b}})\citenamefont {LeDesma}, \citenamefont {Mehling},\ and\ \citenamefont {Holland}}]{ledesma2024vectoratomaccelerometryoptical}%
  \BibitemOpen
  \bibfield  {author} {\bibinfo {author} {\bibfnamefont {C.}~\bibnamefont {LeDesma}}, \bibinfo {author} {\bibfnamefont {K.}~\bibnamefont {Mehling}},\ and\ \bibinfo {author} {\bibfnamefont {M.}~\bibnamefont {Holland}},\ }\href {https://arxiv.org/abs/2407.04874} {\bibinfo {title} {Vector atom accelerometry in an optical lattice}} (\bibinfo {year} {2024}{\natexlab{b}}),\ \Eprint {https://arxiv.org/abs/2407.04874} {arXiv:2407.04874 [quant-ph]} \BibitemShut {NoStop}%
\bibitem [{\citenamefont {Kapitza}\ and\ \citenamefont {Dirac}(1933)}]{Kapitza_Dirac_1933}%
  \BibitemOpen
  \bibfield  {author} {\bibinfo {author} {\bibfnamefont {P.~L.}\ \bibnamefont {Kapitza}}\ and\ \bibinfo {author} {\bibfnamefont {P.~A.~M.}\ \bibnamefont {Dirac}},\ }\bibfield  {title} {\bibinfo {title} {The reflection of electrons from standing light waves},\ }\href {https://doi.org/10.1017/S0305004100011105} {\bibfield  {journal} {\bibinfo  {journal} {Mathematical Proceedings of the Cambridge Philosophical Society}\ }\textbf {\bibinfo {volume} {29}},\ \bibinfo {pages} {297–300} (\bibinfo {year} {1933})}\BibitemShut {NoStop}%
\bibitem [{\citenamefont {Shao}\ \emph {et~al.}(2022)\citenamefont {Shao}, \citenamefont {Combes}, \citenamefont {Hauser},\ and\ \citenamefont {Nicotra}}]{PhysRevA.105.032605}%
  \BibitemOpen
  \bibfield  {author} {\bibinfo {author} {\bibfnamefont {J.}~\bibnamefont {Shao}}, \bibinfo {author} {\bibfnamefont {J.}~\bibnamefont {Combes}}, \bibinfo {author} {\bibfnamefont {J.}~\bibnamefont {Hauser}},\ and\ \bibinfo {author} {\bibfnamefont {M.~M.}\ \bibnamefont {Nicotra}},\ }\bibfield  {title} {\bibinfo {title} {Projection-operator-based newton method for the trajectory optimization of closed quantum systems},\ }\href {https://doi.org/10.1103/PhysRevA.105.032605} {\bibfield  {journal} {\bibinfo  {journal} {Phys. Rev. A}\ }\textbf {\bibinfo {volume} {105}},\ \bibinfo {pages} {032605} (\bibinfo {year} {2022})}\BibitemShut {NoStop}%
\bibitem [{\citenamefont {Shao}\ \emph {et~al.}(2024)\citenamefont {Shao}, \citenamefont {Naris}, \citenamefont {Hauser},\ and\ \citenamefont {Nicotra}}]{PhysRevA.109.012609}%
  \BibitemOpen
  \bibfield  {author} {\bibinfo {author} {\bibfnamefont {J.}~\bibnamefont {Shao}}, \bibinfo {author} {\bibfnamefont {M.}~\bibnamefont {Naris}}, \bibinfo {author} {\bibfnamefont {J.}~\bibnamefont {Hauser}},\ and\ \bibinfo {author} {\bibfnamefont {M.~M.}\ \bibnamefont {Nicotra}},\ }\bibfield  {title} {\bibinfo {title} {Solving quantum optimal control problems using projection-operator-based newton steps},\ }\href {https://doi.org/10.1103/PhysRevA.109.012609} {\bibfield  {journal} {\bibinfo  {journal} {Phys. Rev. A}\ }\textbf {\bibinfo {volume} {109}},\ \bibinfo {pages} {012609} (\bibinfo {year} {2024})}\BibitemShut {NoStop}%
\bibitem [{\citenamefont {Chih}(2022)}]{liang2022thesis}%
  \BibitemOpen
  \bibfield  {author} {\bibinfo {author} {\bibfnamefont {L.-Y.}\ \bibnamefont {Chih}},\ }\emph {\bibinfo {title} {Machine-Learning-Based Design of Quantum Systems for Extreme Sensing}},\ \href@noop {} {Ph.D. thesis},\ \bibinfo {address} {Boulder} (\bibinfo {year} {2022})\BibitemShut {NoStop}%
\bibitem [{\citenamefont {Shao}\ \emph {et~al.}(2023)\citenamefont {Shao}, \citenamefont {Chih}, \citenamefont {Naris}, \citenamefont {Holland},\ and\ \citenamefont {Nicotra}}]{10156455}%
  \BibitemOpen
  \bibfield  {author} {\bibinfo {author} {\bibfnamefont {J.}~\bibnamefont {Shao}}, \bibinfo {author} {\bibfnamefont {L.-Y.}\ \bibnamefont {Chih}}, \bibinfo {author} {\bibfnamefont {M.}~\bibnamefont {Naris}}, \bibinfo {author} {\bibfnamefont {M.}~\bibnamefont {Holland}},\ and\ \bibinfo {author} {\bibfnamefont {M.~M.}\ \bibnamefont {Nicotra}},\ }\bibfield  {title} {\bibinfo {title} {Application of quantum optimal control to shaken lattice interferometry},\ }in\ \href {https://doi.org/10.23919/ACC55779.2023.10156455} {\emph {\bibinfo {booktitle} {2023 American Control Conference (ACC)}}}\ (\bibinfo {year} {2023})\ pp.\ \bibinfo {pages} {4593--4598}\BibitemShut {NoStop}%
\bibitem [{\citenamefont {Alam}\ \emph {et~al.}(2024)\citenamefont {Alam}, \citenamefont {Colussi}, \citenamefont {Wilson}, \citenamefont {Reilly}, \citenamefont {Perlin},\ and\ \citenamefont {Holland}}]{alam2024robustquantumsensingmultiparameter}%
  \BibitemOpen
  \bibfield  {author} {\bibinfo {author} {\bibfnamefont {S.~S.}\ \bibnamefont {Alam}}, \bibinfo {author} {\bibfnamefont {V.~E.}\ \bibnamefont {Colussi}}, \bibinfo {author} {\bibfnamefont {J.~D.}\ \bibnamefont {Wilson}}, \bibinfo {author} {\bibfnamefont {J.~T.}\ \bibnamefont {Reilly}}, \bibinfo {author} {\bibfnamefont {M.~A.}\ \bibnamefont {Perlin}},\ and\ \bibinfo {author} {\bibfnamefont {M.~J.}\ \bibnamefont {Holland}},\ }\href {https://arxiv.org/abs/2405.07907} {\bibinfo {title} {Robust quantum sensing with multiparameter decorrelation}} (\bibinfo {year} {2024}),\ \Eprint {https://arxiv.org/abs/2405.07907} {arXiv:2405.07907 [quant-ph]} \BibitemShut {NoStop}%
\bibitem [{\citenamefont {Suzuki}\ \emph {et~al.}(2020)\citenamefont {Suzuki}, \citenamefont {Yang},\ and\ \citenamefont {Hayashi}}]{Suzuki_2020}%
  \BibitemOpen
  \bibfield  {author} {\bibinfo {author} {\bibfnamefont {J.}~\bibnamefont {Suzuki}}, \bibinfo {author} {\bibfnamefont {Y.}~\bibnamefont {Yang}},\ and\ \bibinfo {author} {\bibfnamefont {M.}~\bibnamefont {Hayashi}},\ }\bibfield  {title} {\bibinfo {title} {Quantum state estimation with nuisance parameters},\ }\href {https://doi.org/10.1088/1751-8121/ab8b78} {\bibfield  {journal} {\bibinfo  {journal} {Journal of Physics A: Mathematical and Theoretical}\ }\textbf {\bibinfo {volume} {53}},\ \bibinfo {pages} {453001} (\bibinfo {year} {2020})}\BibitemShut {NoStop}%
\bibitem [{\citenamefont {Hamilton}\ \emph {et~al.}(2015{\natexlab{b}})\citenamefont {Hamilton}, \citenamefont {Jaffe}, \citenamefont {Brown}, \citenamefont {Maisenbacher}, \citenamefont {Estey},\ and\ \citenamefont {Muller}}]{PhysRevLett.114.100405}%
  \BibitemOpen
  \bibfield  {author} {\bibinfo {author} {\bibfnamefont {P.}~\bibnamefont {Hamilton}}, \bibinfo {author} {\bibfnamefont {M.}~\bibnamefont {Jaffe}}, \bibinfo {author} {\bibfnamefont {J.~M.}\ \bibnamefont {Brown}}, \bibinfo {author} {\bibfnamefont {L.}~\bibnamefont {Maisenbacher}}, \bibinfo {author} {\bibfnamefont {B.}~\bibnamefont {Estey}},\ and\ \bibinfo {author} {\bibfnamefont {H.}~\bibnamefont {Muller}},\ }\bibfield  {title} {\bibinfo {title} {Atom interferometry in an optical cavity},\ }\href {https://doi.org/10.1103/PhysRevLett.114.100405} {\bibfield  {journal} {\bibinfo  {journal} {Phys. Rev. Lett.}\ }\textbf {\bibinfo {volume} {114}},\ \bibinfo {pages} {100405} (\bibinfo {year} {2015}{\natexlab{b}})}\BibitemShut {NoStop}%
\bibitem [{\citenamefont {Panda}\ \emph {et~al.}(2024)\citenamefont {Panda}, \citenamefont {Tao}, \citenamefont {Ceja}, \citenamefont {Khoury}, \citenamefont {Tino},\ and\ \citenamefont {Muller}}]{Panda2024}%
  \BibitemOpen
  \bibfield  {author} {\bibinfo {author} {\bibfnamefont {C.~D.}\ \bibnamefont {Panda}}, \bibinfo {author} {\bibfnamefont {M.~J.}\ \bibnamefont {Tao}}, \bibinfo {author} {\bibfnamefont {M.}~\bibnamefont {Ceja}}, \bibinfo {author} {\bibfnamefont {J.}~\bibnamefont {Khoury}}, \bibinfo {author} {\bibfnamefont {G.~M.}\ \bibnamefont {Tino}},\ and\ \bibinfo {author} {\bibfnamefont {H.}~\bibnamefont {Muller}},\ }\bibfield  {title} {\bibinfo {title} {Measuring gravitational attraction with a lattice atom interferometer},\ }\href@noop {} {\bibfield  {journal} {\bibinfo  {journal} {Nature}\ } (\bibinfo {year} {2024})}\BibitemShut {NoStop}%
\bibitem [{\citenamefont {Viola}\ \emph {et~al.}(1999)\citenamefont {Viola}, \citenamefont {Knill},\ and\ \citenamefont {Lloyd}}]{PhysRevLett.82.2417}%
  \BibitemOpen
  \bibfield  {author} {\bibinfo {author} {\bibfnamefont {L.}~\bibnamefont {Viola}}, \bibinfo {author} {\bibfnamefont {E.}~\bibnamefont {Knill}},\ and\ \bibinfo {author} {\bibfnamefont {S.}~\bibnamefont {Lloyd}},\ }\bibfield  {title} {\bibinfo {title} {Dynamical decoupling of open quantum systems},\ }\href {https://doi.org/10.1103/PhysRevLett.82.2417} {\bibfield  {journal} {\bibinfo  {journal} {Phys. Rev. Lett.}\ }\textbf {\bibinfo {volume} {82}},\ \bibinfo {pages} {2417} (\bibinfo {year} {1999})}\BibitemShut {NoStop}%
\bibitem [{\citenamefont {Levitt}\ and\ \citenamefont {Freeman}(1981)}]{LEVITT198165}%
  \BibitemOpen
  \bibfield  {author} {\bibinfo {author} {\bibfnamefont {M.~H.}\ \bibnamefont {Levitt}}\ and\ \bibinfo {author} {\bibfnamefont {R.}~\bibnamefont {Freeman}},\ }\bibfield  {title} {\bibinfo {title} {Compensation for pulse imperfections in nmr spin-echo experiments},\ }\href {https://doi.org/https://doi.org/10.1016/0022-2364(81)90082-2} {\bibfield  {journal} {\bibinfo  {journal} {Journal of Magnetic Resonance (1969)}\ }\textbf {\bibinfo {volume} {43}},\ \bibinfo {pages} {65} (\bibinfo {year} {1981})}\BibitemShut {NoStop}%
\bibitem [{\citenamefont {Storey}\ and\ \citenamefont {Cohen-Tannoudji}(1994)}]{storey1994feynman}%
  \BibitemOpen
  \bibfield  {author} {\bibinfo {author} {\bibfnamefont {P.}~\bibnamefont {Storey}}\ and\ \bibinfo {author} {\bibfnamefont {C.}~\bibnamefont {Cohen-Tannoudji}},\ }\bibfield  {title} {\bibinfo {title} {The feynman path integral approach to atomic interferometry. a tutorial},\ }\href@noop {} {\bibfield  {journal} {\bibinfo  {journal} {Journal de Physique II}\ }\textbf {\bibinfo {volume} {4}},\ \bibinfo {pages} {1999} (\bibinfo {year} {1994})}\BibitemShut {NoStop}%
\bibitem [{\citenamefont {Barndorff-Nielsen}\ and\ \citenamefont {Gill}(2000)}]{Barndorff-Nielsen_2000}%
  \BibitemOpen
  \bibfield  {author} {\bibinfo {author} {\bibfnamefont {O.~E.}\ \bibnamefont {Barndorff-Nielsen}}\ and\ \bibinfo {author} {\bibfnamefont {R.~D.}\ \bibnamefont {Gill}},\ }\bibfield  {title} {\bibinfo {title} {Fisher information in quantum statistics},\ }\href {https://doi.org/10.1088/0305-4470/33/24/306} {\bibfield  {journal} {\bibinfo  {journal} {Journal of Physics A: Mathematical and General}\ }\textbf {\bibinfo {volume} {33}},\ \bibinfo {pages} {4481} (\bibinfo {year} {2000})}\BibitemShut {NoStop}%
\bibitem [{\citenamefont {Hall}(2000)}]{PhysRevA.62.012107}%
  \BibitemOpen
  \bibfield  {author} {\bibinfo {author} {\bibfnamefont {M.~J.~W.}\ \bibnamefont {Hall}},\ }\bibfield  {title} {\bibinfo {title} {Quantum properties of classical fisher information},\ }\href {https://doi.org/10.1103/PhysRevA.62.012107} {\bibfield  {journal} {\bibinfo  {journal} {Phys. Rev. A}\ }\textbf {\bibinfo {volume} {62}},\ \bibinfo {pages} {012107} (\bibinfo {year} {2000})}\BibitemShut {NoStop}%
\bibitem [{\citenamefont {Clauser}(1988)}]{CLAUSER1988262}%
  \BibitemOpen
  \bibfield  {author} {\bibinfo {author} {\bibfnamefont {J.~F.}\ \bibnamefont {Clauser}},\ }\bibfield  {title} {\bibinfo {title} {Ultra-high sensitivity accelerometers and gyroscopes using neutral atom matter-wave interferometry},\ }\href {https://doi.org/https://doi.org/10.1016/0378-4363(88)90176-3} {\bibfield  {journal} {\bibinfo  {journal} {Physica B+C}\ }\textbf {\bibinfo {volume} {151}},\ \bibinfo {pages} {262} (\bibinfo {year} {1988})}\BibitemShut {NoStop}%
\bibitem [{\citenamefont {De~Broglie}(1929)}]{de1929wave}%
  \BibitemOpen
  \bibfield  {author} {\bibinfo {author} {\bibfnamefont {L.}~\bibnamefont {De~Broglie}},\ }\bibfield  {title} {\bibinfo {title} {The wave nature of the electron},\ }\href@noop {} {\bibfield  {journal} {\bibinfo  {journal} {Nobel lecture}\ }\textbf {\bibinfo {volume} {12}},\ \bibinfo {pages} {244} (\bibinfo {year} {1929})}\BibitemShut {NoStop}%
\bibitem [{\citenamefont {Schubert}\ \emph {et~al.}(2021)\citenamefont {Schubert}, \citenamefont {Abend}, \citenamefont {Gersemann}, \citenamefont {Gebbe}, \citenamefont {Schlippert}, \citenamefont {Berg},\ and\ \citenamefont {Rasel}}]{Schubert2021}%
  \BibitemOpen
  \bibfield  {author} {\bibinfo {author} {\bibfnamefont {C.}~\bibnamefont {Schubert}}, \bibinfo {author} {\bibfnamefont {S.}~\bibnamefont {Abend}}, \bibinfo {author} {\bibfnamefont {M.}~\bibnamefont {Gersemann}}, \bibinfo {author} {\bibfnamefont {M.}~\bibnamefont {Gebbe}}, \bibinfo {author} {\bibfnamefont {D.}~\bibnamefont {Schlippert}}, \bibinfo {author} {\bibfnamefont {P.}~\bibnamefont {Berg}},\ and\ \bibinfo {author} {\bibfnamefont {E.~M.}\ \bibnamefont {Rasel}},\ }\bibfield  {title} {\bibinfo {title} {Multi-loop atomic sagnac interferometry},\ }\href {https://doi.org/10.1038/s41598-021-95334-7} {\bibfield  {journal} {\bibinfo  {journal} {Scientific Reports}\ }\textbf {\bibinfo {volume} {11}},\ \bibinfo {pages} {16121} (\bibinfo {year} {2021})}\BibitemShut {NoStop}%
\bibitem [{\citenamefont {Browaeys}\ \emph {et~al.}(2005)\citenamefont {Browaeys}, \citenamefont {Haffner}, \citenamefont {McKenzie}, \citenamefont {Rolston}, \citenamefont {Helmerson},\ and\ \citenamefont {Phillips}}]{PhysRevA.72.053605}%
  \BibitemOpen
  \bibfield  {author} {\bibinfo {author} {\bibfnamefont {A.}~\bibnamefont {Browaeys}}, \bibinfo {author} {\bibfnamefont {H.}~\bibnamefont {Haffner}}, \bibinfo {author} {\bibfnamefont {C.}~\bibnamefont {McKenzie}}, \bibinfo {author} {\bibfnamefont {S.~L.}\ \bibnamefont {Rolston}}, \bibinfo {author} {\bibfnamefont {K.}~\bibnamefont {Helmerson}},\ and\ \bibinfo {author} {\bibfnamefont {W.~D.}\ \bibnamefont {Phillips}},\ }\bibfield  {title} {\bibinfo {title} {Transport of atoms in a quantum conveyor belt},\ }\href {https://doi.org/10.1103/PhysRevA.72.053605} {\bibfield  {journal} {\bibinfo  {journal} {Phys. Rev. A}\ }\textbf {\bibinfo {volume} {72}},\ \bibinfo {pages} {053605} (\bibinfo {year} {2005})}\BibitemShut {NoStop}%
\end{thebibliography}
\providecommand{\noopsort}[1]{}\providecommand{\singleletter}[1]{#1}%

\end{document}